\documentclass[10pt,twocolumn]{article}

\usepackage[T1]{fontenc}
\usepackage{lmodern}

\usepackage{amsmath,amssymb,amsthm}

\usepackage{graphicx}
\usepackage{float}
\usepackage{booktabs}
\usepackage{multirow}
\usepackage{tabularx}
\usepackage{array}
\newcolumntype{Y}{>{\raggedright\arraybackslash}Y}
\newcolumntype{Z}{>{\centering\arraybackslash}Z}
\renewcommand{\arraystretch}{1.3}
\usepackage{subcaption}

\usepackage{pifont}
\newcommand{\cmark}{\ding{51}}
\newcommand{\xmark}{\ding{55}}

\usepackage{enumitem}
\setlist{topsep=3pt, itemsep=3pt, parsep=0pt, partopsep=0pt}
\setlist[enumerate,1]{label=\alph*), leftmargin=*}

\usepackage[numbers,sort&compress]{natbib}

\usepackage[colorlinks=true,linkcolor=blue,citecolor=blue,urlcolor=blue]{hyperref}

\usepackage{geometry}
\newtheorem{proposition}{Proposition}

\usepackage{titlesec}
\titleformat{\section}{\normalsize\bfseries}{\thesection.}{0.5em}{}
\titlespacing*{\section}{0pt}{2.5ex plus 0.2ex minus 0.2ex}{1.0ex}
\titleformat{\subsection}{\normalsize\itshape}{\thesubsection}{0.5em}{}
\titlespacing*{\subsection}{0pt}{1.5ex plus 0.2ex minus 0.2ex}{0.8ex}
\titleformat{\subsubsection}{\normalsize\itshape}{\thesubsubsection}{0.6em}{}
\titlespacing*{\subsubsection}{0pt}{1.2ex plus 0.2ex minus 0.2ex}{0.6ex}
\titleformat{\paragraph}{\normalsize\bfseries}{}{0em}{}
\titlespacing*{\paragraph}{0pt}{1.2ex}{0.2em}

\usepackage[font=small,labelfont=bf]{caption}
\DeclareMathOperator{\logit}{logit}

\newcommand{\USL}{\mathrm{USL}}
\newcommand{\LSL}{\mathrm{LSL}}

\usepackage{threeparttable}

\usepackage[dvipsnames]{xcolor}
\usepackage[normalem]{ulem}
\usepackage{soul}
\sethlcolor{yellow!35}

\usepackage{framed}

\usepackage{fancyhdr}
\title{A Hybrid Statistical Learning Framework for Capability-Based Decision Support in Manufacturing}

\author{
	Fei Jiang\thanks{
		Independent Researcher, Seattle, WA, USA.
		Corresponding author: Fei Jiang.
		E-mail: jiangfeicq@gmail.com.
		Fei Jiang and Lei Yang contributed equally to this work.
	}
	\and
	Lei Yang\footnotemark[1]
}

\date{}
\begin{document}
	\maketitle
	\thispagestyle{fancy}
	\pagestyle{plain} 
	
\begin{abstract}
	Process capability indices are widely used in manufacturing quality control, but capability approval is often implemented by directly thresholding finite-sample estimates, which can produce unstable and poorly calibrated decisions near the approval boundary. This paper develops a hybrid statistical--learning framework for capability-based decision support in manufacturing. The proposed UC-Cap approach combines a statistically grounded capability baseline with a residual learning component that uses process, distributional, specification-related, and measurement-related features to refine capability-decision risk estimates under non-ideal manufacturing conditions. The statistical baseline preserves the interpretability of classical capability analysis, while the learning component provides data-driven correction for systematic deviations arising from non-normality, measurement effects, and finite-sample variability. A nested Monte Carlo evaluation is introduced to assess probabilistic calibration under controlled synthetic settings, and an empirical manufacturing study is used to examine decision behavior under realistic capability data. Results show that deterministic thresholding can lead to substantial instability and miscalibration in near-boundary regimes, whereas the proposed framework provides calibrated risk estimates, interpretable decision outputs, and improved support for capability approval decisions. The framework is compatible with existing capability-analysis workflows and can be integrated into manufacturing quality decision-support systems.
\end{abstract}

	
\section{Introduction}
\label{sec:introduction}

Process capability indices (PCIs) such as $C_{pk}$ have long served as fundamental tools for assessing process performance and supporting quality-related decision making in manufacturing systems \cite{kane1986process, kotz2002process, montgomery2020introduction}. These indices provide a concise summary of how well a process conforms to specification limits, and are widely embedded in industrial standards and guidelines for statistical process control and capability analysis \cite{ISO22514-1-2014, ISO22514-4-2016, oakland2007statistical}. Despite their widespread adoption, capability approval is typically implemented as a deterministic threshold rule applied to finite-sample estimates, which ignores estimation variability and can lead to unstable decisions near the capability boundary.

In practice, capability approval is typically defined as a deterministic rule of the form $\widehat{C}_{pk} \ge C_0$, where $\widehat{C}_{pk}$ is estimated from finite samples and $C_0$ is a prescribed threshold. This rule is simple, interpretable, and widely adopted in engineering practice \cite{jiang2026practical}. 
However, it implicitly assumes that the estimated capability is a reliable proxy for the true process capability, thereby neglecting the uncertainty inherent in finite-sample estimation.

A substantial body of literature has studied statistical properties of capability indices, including distributional behavior, estimation accuracy, and sensitivity to assumptions \cite{pearn1992distributional, bissell1990reliable, mahmoud2010estimating}. Extensions have also been developed to address non-normal distributions \cite{chen1997application, clements1989process}, asymmetric tolerances \cite{chen2001capability, abbasi2016class}, and alternative formulations such as $C_{pm}$ and related indices \cite{chan1988new, boyles1991taguchi, vannman1995unified}. These works significantly enrich theoretical and practical understanding of capability indices, yet largely focus on point or interval estimation, rather than reliability of decision-making based on these estimates.

One critical but underexplored issue is the effect of finite-sample uncertainty on capability-based decisions. 
When the true capability lies close to the threshold $C_0$, the estimator $\widehat{C}_{pk}$ exhibits non-negligible variability, and repeated sampling may cause the estimate to cross the decision boundary. 
This leads to substantial decision inconsistency, where identical processes may be alternately accepted or rejected depending on sampling variation. 

Recent work has formally characterized this phenomenon as finite-sample decision instability, demonstrating that misclassification risk is inherently non-negligible near the capability boundary \cite{jiang2026finite}. 
Related studies in measurement uncertainty and conformity assessment have also emphasized that decision rules ignoring uncertainty can lead to systematically biased or unreliable outcomes \cite{pendrill2014using, ISO14253-1-2013, desimoni2011uncertainty}.

These observations suggest that the classical deterministic rule is fundamentally incomplete. 
A more appropriate formulation is to treat capability approval as a decision problem under uncertainty, where the objective is to estimate decision risk under finite-sample variability in a statistically consistent manner.
This perspective is naturally aligned with statistical decision theory \cite{wald1950statistical, degroot2005optimal, berger2013statistical}, in which optimal decisions are derived by balancing different types of risks.

Recent advances in industrial analytics and data-driven modeling have enabled more flexible approaches for analyzing complex manufacturing and quality-related decision problems, including methods based on logistic regression \cite{hosmer2013applied} and gradient boosting \cite{chen2016xgboost}. However, when applied to capability approval, purely data-driven approaches often treat the problem as a generic classification task, without explicitly incorporating the statistical structure of capability estimation or the effects of finite-sample uncertainty. As a result, predicted probabilities may become unstable or insufficiently calibrated in decision-critical regions, limiting their reliability for uncertainty-sensitive manufacturing decisions \cite{gneiting2007strictly, guo2017calibration, senge2014reliable}.

In this work, we propose UC-Cap (Uncertainty-Calibrated Capability), an uncertainty-aware framework for capability-based decision making under finite-sample variability. The framework combines a statistically grounded baseline, representing the uncertainty-adjusted distance to the capability threshold, with an adaptive correction component that accounts for systematic deviations arising from non-normality, measurement effects, and other non-ideal process conditions. This formulation provides an uncertainty-aware estimate of decision risk, particularly in near-threshold regions where approval decisions are most sensitive to sampling variability, while preserving interpretability through its structured capability-based formulation. Building on recent advances in capability analysis and risk-based decision modeling \cite{jiang2026finite,jiang2026risk}, UC-Cap extends classical capability analysis into a probabilistic decision-support framework for manufacturing quality decisions.

The main contributions of this work are as follows:
\begin{enumerate}
	\item We formulate capability approval as a capability-based decision-support problem in manufacturing, replacing direct deterministic thresholding with calibrated risk estimation from finite-sample capability evidence.

	\item We show that near-boundary decision behavior is jointly governed by capability margin, estimation variability, and process-specific characteristics, explaining why classical threshold rules can become unstable in borderline cases.
	
	\item We propose UC-Cap, a hybrid statistical--learning framework that combines a capability-based statistical baseline with residual learning from distributional, specification-related, and measurement-related features.
	
	\item We introduce a nested Monte Carlo evaluation procedure for assessing calibration and decision-risk estimation under controlled synthetic manufacturing scenarios.
	
	\item We demonstrate the practical value of the proposed framework through simulation and empirical manufacturing data, showing how calibrated risk scores can support interpretable capability approval decisions.
\end{enumerate}

The remainder of the paper is organized as follows.
Sections~\ref{sec:stat_foundations}--\ref{sec:decision_risk} establish the statistical formulation and uncertainty-aware capability decision framework.
Sections~\ref{sec:training_decision}--\ref{dimetra} present the model implementation and engineering decision workflow.
Section~\ref{sec:simulation_evaluation} provides simulation-based evaluation under controlled conditions, while Section~\ref{sec:empirical_validation} presents empirical validation on manufacturing data.
Finally, Section~\ref{sec:conclusion} concludes the paper.

\section{Statistical Foundations of Capability Analysis}
\label{sec:stat_foundations}

In this paper, $C_{pk}$ is used as a generic notation for capability indices. 
The proposed framework is agnostic to both distributional assumptions and variability estimation: 
while classical definitions assume normality, empirical analyses (Sections~\ref{dimetra}--\ref{sec:empirical_validation}) 
use overall standard deviation and adopt percentile-based estimation when normality is violated, 
without changing the underlying probabilistic formulation.

\subsection{Process Capability and Classical Approval Rule}

In manufacturing quality control, process capability is commonly quantified using the index \cite{kotz2002process}
\begin{equation}
	C_{pk} = \min\left( \frac{\USL - \mu}{3\sigma}, \frac{\mu - \LSL}{3\sigma} \right),
\end{equation}
which assumes normally distributed outputs with bilateral tolerances, where $\USL$ and $\LSL$ are the upper and lower specification limits, $\mu$ and $\sigma$ denote the process mean and standard deviation.

In practice, the true (population-level) capability $C_{pk}^{\mathrm{true}}$ is unknown and estimated from finite samples as $\widehat C_{pk}$. A widely adopted approval rule is the fixed-threshold criterion \cite{ISO22514-1-2014,ISO22514-4-2016}:
\begin{equation}
	\widehat C_{pk} \ge C_0,
\end{equation}
where $C_0$ is a predefined capability threshold.

This rule treats $\widehat C_{pk}$ as a proxy for $C_{pk}^{\mathrm{true}}$; however, under finite-sample conditions, this approximation may be unreliable, leading to substantial decision uncertainty.

This perspective is consistent with established conformity assessment practices in metrology, where decision rules explicitly incorporate measurement uncertainty. For example, guard-banding approaches in JCGM 106 \cite{jcgm106} adjust acceptance thresholds to control the risks of false acceptance and false rejection, and ISO/IEC 17025 emphasizes the role of uncertainty in compliance decisions \cite{iso17025}.

Within this context, the statistical baseline
\begin{equation}
	z^{(stat)} = \frac{C_0 - \widehat{C}_{pk}}{SE}
	\label{eq:z_stat_signal_noise}
\end{equation}
can be interpreted as a normalized signal-to-noise ratio governing decision uncertainty near the capability threshold, and as a continuous analog of guard-banding, where the normalized distance to the threshold reflects both estimation and uncertainty. Unlike classical fixed-margin or worst-case approaches, this formulation is adaptive under finite-sample uncertainty.

While the statistical baseline is derived under a probit link, the residual component operates on the log-odds scale. This choice is motivated by the local equivalence between probit and logit links near the decision boundary, where both transformations are approximately linear. In particular,
\[
\Phi(z) \approx \sigma(kz), \quad k \approx \pi/\sqrt{3},
\]
which implies that additive corrections in the log-odds space provide a flexible and numerically stable approximation to deviations from the probit-based baseline.

Building on this interpretation, the proposed model learns an uncertainty-aware decision function that adapts to both estimation variability and feature-dependent structure. It can also be viewed as a frequentist counterpart to posterior-based acceptance criteria in Bayesian conformity assessment, where decisions are based on the probability of compliance given uncertainty.

\subsection{Boundary Instability of Fixed-Threshold Decisions}
A fundamental limitation of the fixed-threshold rule arises near the capability boundary. When $C_{pk}^{\mathrm{true}} = C_0$, the acceptance probability satisfies
\begin{equation}
	P(\widehat C_{pk} \ge C_0 \mid C_{pk}^{\mathrm{true}} = C_0) \approx 0.5,
\end{equation}
indicating that the decision becomes effectively random at the boundary \cite{jiang2026finite}. More generally, when $C_{pk}^{\mathrm{true}}$ lies near $C_0$, the acceptance probability deviates from 0 or 1, implying substantial decision uncertainty. This shows that capability approval near the threshold is inherently probabilistic rather than deterministic, motivating a risk-based formulation.

\subsection{Failure Probability Formulation under Finite Samples}
\label{subsec:failure_probability}

We model capability approval as a probabilistic decision problem. For each dimension $j$, define failure probability.
\begin{equation}
	\pi_j = P(C_{pk,j}^{\mathrm{true}} < C_0 \mid D_j),
\end{equation}
where $D_j$ denotes the observed data.

Under standard asymptotic approximations \cite{van2000asymptotic,serfling2009approximation},
\begin{equation}
	\widehat C_{pk,j}\mid C_{pk,j}^{\mathrm{true}}\approx
	\mathcal N\!\left(C_{pk,j}^{\mathrm{true}},\,SE(\widehat C_{pk,j})^2\right),
	\label{eq:cpk_sampling_normal}
\end{equation}
which yields
\begin{equation}
	\pi_j \approx
	\Phi\left(
	\frac{C_0 - \widehat C_{pk,j}}{SE_j}
	\right),
	\label{eq:failure_prob_approx}
\end{equation}
where $\Phi(\cdot)$ is the standard normal CDF and $SE_j$ is the standard error of $\widehat C_{pk,j}$.

This formulation transforms capability evaluation into a risk quantification problem under finite-sample uncertainty \cite{jiang2026risk}. The resulting probability should therefore be interpreted as a finite-sample decision-risk approximation rather than a direct physical defect-generation probability.

\subsection{Decision-Theoretic Perspective}
\label{subsec:decision_perspective}

Given the failure probability $\pi_j$, capability approval can be interpreted as a decision problem under uncertainty \cite{degroot2005optimal}. Let $c_{FA}$ and $c_{FR}$ denote the costs of false acceptance and false rejection. The corresponding Bayes-optimal rule is \cite{wald1950statistical}
\begin{equation}
	\text{approve if } \pi_j \le \alpha,
	\quad
	\alpha = \frac{c_{FR}}{c_{FA} + c_{FR}}.
	\label{eq:bayes_decision_rule}
\end{equation}

This formulation replaces deterministic thresholding with a risk-based rule in which the decision boundary is determined by relative costs and explicitly accounts for uncertainty. In this sense, classical threshold-based and guard-band rules can be viewed as implicit approximations to such risk-based decision boundaries.

In the proposed framework, this perspective serves as an interpretation rather than an optimization objective. The UC-Cap model focuses on learning a calibrated estimate of $\pi_j$, which can be mapped to application-specific decision thresholds depending on cost considerations. This provides a flexible link between statistical modeling and decision-making without requiring explicit cost-sensitive training in the current implementation.

\subsection{Limitations of Existing Approaches}

The approximation in \eqref{eq:failure_prob_approx}, denoted as \(\pi_j^{(stat)}\), provides a principled baseline for failure risk. However, it relies on asymptotic normality and may be inaccurate under small sample sizes or non-normal process distributions \cite{chen1997application}.

Moreover, this formulation depends primarily on $\widehat C_{pk}$ and $SE_j$, and thus captures only limited information about the underlying process. It does not account for factors such as distributional shape, specification geometry, or other process-specific characteristics \cite{jiang2026practical,deleryd1998gap}. 

In contrast, purely data-driven approaches (e.g., logistic regression or gradient boosting) can model nonlinear relationships and feature interactions, but typically lack explicit incorporation of statistical structure and uncertainty in capability estimation \cite{hosmer2013applied,chen2016xgboost}. These limitations are complementary: statistical approaches provide structure but limited flexibility, while data-driven models offer flexibility but lack principled uncertainty awareness.

To address this gap, we propose a hybrid framework that integrates a statistically grounded baseline with a data-driven residual correction, yielding a theory-informed and uncertainty-aware modeling approach.

We note that alternative uncertainty-aware approaches, such as Bayesian hierarchical models and conformal prediction methods, provide complementary perspectives on decision calibration. While these methods offer stronger theoretical guarantees, they typically involve more complex modeling assumptions or additional computational overhead, which is beyond the scope of this work; a detailed comparison is left for future work.

\section{Decision-Risk and UC-Cap Framework}
\label{sec:decision_risk}

Capability approval decisions can become highly unstable under finite-sample conditions, particularly near the acceptance threshold, where estimation variability directly translates into decision risk. To address this, we develop an uncertainty-aware probabilistic framework that explicitly models and corrects this instability through a combination of statistical structure and data-driven learning.

\subsection{Statistical Baseline Modeling}
\label{subsec:stat_baseline}

Under standard regularity conditions, the estimator $\widehat{C}_{pk,j}$ follows the asymptotic normal approximation in \eqref{eq:cpk_sampling_normal}. Using a confidence-distribution argument \cite{lehmann1998theory,casella2024statistical}, we obtain
\begin{equation}
	C_{pk}^{\mathrm{true}} \mid D
	\approx
	\mathcal{N}\!\left(\widehat{C}_{pk},\, SE(\widehat{C}_{pk})^2\right),
	\label{eq:posterior_approx}
\end{equation}
which leads to the failure probability approximation
\begin{equation}
	P(C_{pk}^{\mathrm{true}}<C_0 \mid D)
	\approx
	\Phi\!\left(
	\frac{C_0 - \widehat{C}_{pk}}{SE(\widehat{C}_{pk})}
	\right).
	\label{eq:pfail_stat}
\end{equation}

For dimension $j$, we define the statistical baseline
\begin{equation}
	\pi_j^{(stat)} =
	\Phi\!\left(
	\frac{C_0 - \widehat{C}_{pk,j}}{SE_j}
	\right),
	\quad
	SE_j := SE(\widehat{C}_{pk,j}),
	\label{eq:pi_stat}
\end{equation}
which provides a confidence-based approximation of the probability that true capability falls below the threshold.

This formulation relies on asymptotic normality and may degrade under small samples or non-normality, particularly near the decision boundary where estimation errors strongly affect decisions. In addition, the non-smooth min operator in $C_{pk}$ can further induce deviations from normality in finite samples. These limitations motivate the residual correction introduced in Section~\ref{uc-cap}.

The behavior of $\pi_j^{(stat)}$ can be summarized in three regimes: (i) $\pi_j^{(stat)} \approx 0$ when $\widehat{C}_{pk,j} \gg C_0$ (clearly capable), (ii) $\pi_j^{(stat)} \approx 0.5$ when $\widehat{C}_{pk,j} \approx C_0$ (boundary uncertainty), and (iii) moderate $\pi_j^{(stat)}$ when $\widehat{C}_{pk,j} < C_0$ but $SE_j$ is large (uncertain sub-threshold). These regimes highlight that $\pi_j^{(stat)}$ provides a smooth, uncertainty-aware alternative to deterministic thresholding.

\subsection{Structural Decomposition of Decision Risk}
\label{uc-cap}
The approximation in Section~\ref{subsec:stat_baseline} indicates that capability approval under finite samples is inherently probabilistic, rather than a deterministic thresholding problem. Decision reliability depends on the relative position of the capability estimate with respect to the approval threshold, normalized by its estimation uncertainty.

This motivates a signal-to-noise representation of decision risk. Let $C$ denote the latent (true) process capability and $U$ the associated estimation uncertainty.

\begin{proposition}[Structural form of capability decision risk]
	Under standard regularity conditions, the probability of failing the capability requirement can be approximated as
	\begin{equation}
		\pi \approx \Phi\!\left(\frac{C_0 - C}{U}\right),
	\end{equation}
	where $\Phi(\cdot)$ is the standard normal cumulative distribution function.
\end{proposition}

This implies that decision risk is governed by the normalized distance to the threshold,
\[
z = \frac{C_0 - C}{U},
\]
where the capability margin $(C - C_0)$ acts as signal and $U$ as noise.

The representation induces structural constraints: for fixed $U$, risk is monotone in the capability margin, while for fixed $C$, increasing uncertainty increases risk. This motivates preserving the statistical baseline defined by $z$, and learning only systematic deviations through a residual component, rather than modeling risk as an unconstrained function of features \cite{hand2006classifier,guo2017calibration}.

\subsection{Hybrid Risk Decomposition}

Motivated by the structural form in Section~\ref{uc-cap}, we formulate UC-Cap as a theory-anchored residual model that refines a statistically grounded baseline for capability decision risk.

For model implementation, the dimension-level statistical baseline is represented on the log-odds scale. Let $x_j$ denote the feature vector for dimension $j$. Then
\begin{equation}
	z_j^{(stat)} = \logit\bigl(\pi_j^{(stat)}\bigr)
	= \log\left( \frac{\pi_j^{(stat)}}{1 - \pi_j^{(stat)}} \right),
	\label{eq:z_stat}
\end{equation}
as the log-odds representation of the statistical baseline.

The UC-Cap model is
\begin{equation}
	\logit(\pi_j) = z_j^{(stat)} + f_\theta(x_j),
	\label{eq:logit_residual}
\end{equation}
or equivalently,
\begin{equation}
	\pi_j = \sigma\!\left(z_j^{(stat)} + f_\theta(x_j)\right),
	\label{eq:residual_main}
\end{equation}
where $f_\theta(x_j)$ is a learnable residual function and $\sigma(t)=1/(1+e^{-t})$.

The baseline encodes the uncertainty-adjusted distance to the capability threshold, while $f_\theta(x_j)$ captures systematic deviations due to non-normality, measurement effects, and higher-order feature interactions. When $f_\theta(x_j)\equiv 0$, the model reduces to the statistical baseline.

\subsection{Risk Representation and Decision Interpretation}
The log-odds representation in \eqref{eq:z_stat} enables additive integration of the statistical baseline and the data-driven residual in \eqref{eq:logit_residual}. Although the baseline originates from a probit approximation, the logit scale provides a convenient and numerically stable representation. The two links are locally equivalent near the decision boundary, where most decision uncertainty is concentrated, so the additive log-odds formulation provides an accurate and flexible approximation in the regime of interest.

Traditional capability approval is a deterministic rule:
\begin{equation}
	a_j = \mathbb{I}(\widehat{C}_{pk,j} \ge C_0),
	\label{eq:deterministic_decision_rule}
\end{equation}
which ignores estimation uncertainty. In contrast, UC-Cap induces a risk-based decision rule \cite{degroot2005optimal}:
\begin{equation}
	a_j = \mathbb{I}(\pi_j \le \alpha),
	\label{eq:risk_based_decision_rule}
\end{equation}
where $\pi_j$ is the failure probability and $\alpha$ is a user-specified tolerance, enabling uncertainty-aware decision making.

The statistical baseline $\pi_j^{(stat)}$ captures the dominant effect of finite-sample uncertainty under idealized assumptions, but may deviate from real-world behavior due to non-normality, measurement variability, and feature interactions \cite{chen2001capability}. Under the formulation in \eqref{eq:logit_residual}, the residual term $f_\theta(x_j)$ captures these systematic deviations in a data-driven manner, while the baseline remains the primary signal.

This yields a structured calibration model that preserves interpretability while improving robustness, particularly near the decision boundary where statistical approximations are most fragile \cite{breiman1996stacked,hastie2017generalized}.
\section{UC-Cap Model Formulation, Training, and Decision Framework}
\label{sec:training_decision}

\subsection{Training Objective Formulation}
\label{subsec:training_objectives}

Suppose binary targets $y_j \in \{0,1\}$ are available. 
The model can be trained using the Bernoulli negative log-likelihood (binary cross-entropy, BCE) \cite{hosmer2013applied}:
\begin{equation}
	\mathcal{L}_{\mathrm{BCE}}
	=
	- \sum_{j=1}^J
	\left[
	y_j \log \pi_j
	+
	(1 - y_j)\log(1 - \pi_j)
	\right].
	\label{eq:bce_loss}
\end{equation}

When probabilistic targets $\widetilde y_j \in [0,1]$ are available (e.g., from bootstrap or analytical estimates), the same objective extends naturally to soft cross-entropy:
\begin{equation}
	\mathcal{L}_{\mathrm{soft\text{-}CE}}
	=
	- \sum_{j=1}^J
	\left[
	\widetilde y_j \log \pi_j
	+
	(1 - \widetilde y_j)\log(1 - \pi_j)
	\right],
	\label{eq:soft_ce}
\end{equation}
or alternatively a Brier loss:
\begin{equation}
	\mathcal{L}_{\mathrm{Brier}}
	=
	\sum_{j=1}^J (\pi_j - \widetilde y_j)^2.
	\label{eq:brier_loss}
\end{equation}

In the empirical setting considered in this paper, such soft targets are interpreted as decision-risk surrogates rather than fully independent ground-truth probabilities, since they are partially derived from the same dimension-level statistical summaries used in feature construction. Their primary role is to soften the degeneracy of hard threshold labels and provide smoother supervision for uncertainty-aware capability modeling.

To reduce information leakage during empirical evaluation, the implementation adopts a split-sample construction strategy whenever soft targets are generated from resampling procedures.

Regularization can be incorporated to control model complexity. For example, for a linear residual model:
\begin{equation}
	\mathcal{L}
	=
	\mathcal{L}_{\mathrm{BCE}}
	+
	\lambda_2 \|\beta\|_2^2,
	\label{eq:l2_regularized_loss}
\end{equation}
where $\lambda_2 > 0$ is a regularization parameter controlling the magnitude of residual coefficients, which penalizes large residual coefficients and preserves the dominance of the statistical baseline.

\subsection{Decision Interpretation and Relation to Guard-Band Methods}
\label{subsec:decision_layer}

The model outputs an estimated decision risk $\pi_j$, which can be translated into an approval decision using the risk-based rule in \eqref{eq:risk_based_decision_rule}. 
Unlike deterministic thresholding, the decision threshold is not fixed during training and may instead be selected according to application-specific risk tolerance, deployment requirements, or operational cost considerations. 
This separation between probability estimation and operational decision making enables flexible deployment across different manufacturing and quality-control settings without retraining the underlying model.

The proposed statistical baseline is structurally related to classical guard-band rules used in capability-based conformity assessment, where acceptance thresholds are adjusted according to estimation uncertainty. 
Under the probabilistic formulation $\pi_j^{(stat)} \le \alpha$, the resulting decision boundary naturally incorporates uncertainty information through a continuous risk representation rather than a fixed deterministic margin.

Within the UC-Cap framework, this guard-band structure is preserved through the statistical baseline and further refined through data-driven residual correction under non-ideal manufacturing conditions \cite{pendrill2014using,ISO14253-1-2013}. 
Under this interpretation, UC-Cap generalizes deterministic guard-band approval into a probabilistic uncertainty-aware decision framework, allowing capability approval decisions to adapt continuously to finite-sample uncertainty rather than relying solely on hard thresholding.
\subsection{UC-Cap Model Structure and Residual Learning}

The implemented UC-Cap model follows the additive log-odds residual formulation introduced in Section~\ref{sec:decision_risk}. 
The statistical baseline provides the primary uncertainty-aware capability signal, while the residual component captures systematic deviations associated with non-normality, measurement effects, and higher-order distributional structure. 
Unless otherwise specified, all experiments use the anchored residual formulation with soft-label supervision.

Here, $z_j^{(stat)}$ denotes the baseline log-odds derived from statistical capability theory, while $f_\theta(x_j)$ represents the residual correction component. 
The baseline encodes the dominant uncertainty-adjusted capability signal, whereas the residual captures systematic deviations arising from non-normality, distributional asymmetry, measurement effects, and higher-order distributional characteristics. 
Under the additive residual formulation, the statistical baseline remains the dominant monotonic component, while the residual provides localized refinement when asymptotic assumptions become inaccurate.

For each dimension $j$, the statistical baseline probability $\pi_j^{(stat)}$ is computed according to \eqref{eq:pi_stat} and transformed into the log-odds representation defined in \eqref{eq:z_stat}. 
To ensure numerical stability near the probability boundaries, clipping is applied:
\begin{equation}
	\pi_j^{(stat)} \leftarrow \min\{\max(\pi_j^{(stat)},\epsilon),\, 1-\epsilon\},
	\label{eq:pi_clipping}
\end{equation}
where $\epsilon$ is a small constant (e.g., $10^{-6}$). 
The clipped probability is then transformed to the log-odds scale using \eqref{eq:z_stat}.

The feature vector $x_j$ captures sources of variation not fully represented by $\widehat{C}_{pk,j}$ and $SE_j$. 
These include:

(i) distributional characteristics, such as normality diagnostics, skewness, and kurtosis;

(ii) specification-related structure, including specification type, specification width, distance-to-specification measures, and mean shift; and

(iii) measurement and workflow effects, including measurement-system proxies \cite{aiag2010msa} and analysis-path indicators.

These variables allow the residual component to learn systematic deviations from the assumptions underlying the statistical baseline, particularly under non-normality, asymmetric tolerancing, and measurement uncertainty.

The residual component $f_\theta(x)$ is implemented as a linear model on standardized features with an L2 penalty, ensuring controlled model capacity and stable optimization. 
No explicit interaction terms or nonlinear feature expansions are included, preserving interpretability and reducing overfitting. The regularization strength is selected via validation.

To maintain the anchored structure, the feature set excludes direct transformations of the baseline statistics (e.g., $\widehat{C}_{pk}$ and $SE$), ensuring that $z^{(stat)}$ remains the dominant signal. 
Under this design, the residual captures only systematic deviations beyond the primary statistical structure rather than overriding the underlying capability-based relationship.

To further preserve the structural behavior implied by the statistical baseline, the framework restricts residual capacity through L2 regularization and constrained feature design. Under this formulation, the residual primarily acts as a second-order correction layer, while the dominant relationship between uncertainty and decision risk remains governed by the statistical baseline.

This constrained residual structure enables data-driven refinement while preserving the interpretability and uncertainty-consistent behavior of the underlying capability model. 
Empirically, the residual component mainly contributes near the capability boundary, where asymptotic approximations become less reliable, while its effect remains limited in clearly separable regimes.
\section{Manufacturing Decision-Support Workflow}
\label{dimetra}
\begin{figure}[!t]
	\centering
	\includegraphics[width=1.00\linewidth]{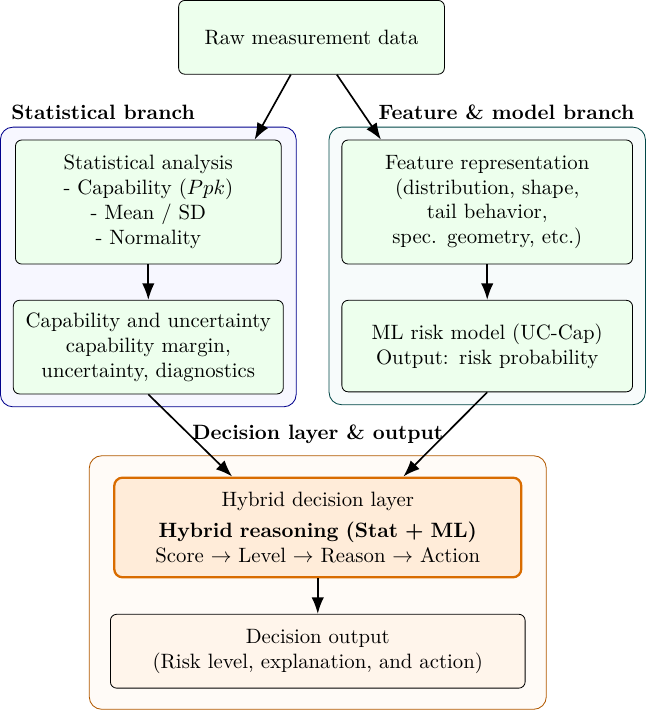}
	\caption{
		System workflow for integrating the proposed UC-Cap framework into a manufacturing quality decision-support pipeline.
		Raw measurement data are processed through two complementary branches: a statistical branch summarizing capability and uncertainty, and a feature/model branch producing data-driven risk estimates.
		These signals are fused in a hybrid decision layer to generate an interpretable decision chain \((\text{score} \rightarrow \text{level} \rightarrow \text{reason} \rightarrow \text{action})\).
	}
	\label{fig:dimetra_workflow}
\end{figure}

\subsection{System Architecture and Decision Output}
Figure~\ref{fig:dimetra_workflow} presents the system-level implementation of the proposed framework in Dimetra, where UC-Cap is integrated with classical capability analysis into a unified decision pipeline. The system consists of two complementary branches: a statistical branch that summarizes capability-related signals, including capability indices, uncertainty measures, and distributional diagnostics, and a feature/model branch that extracts distributional and specification-related features and produces risk estimates via the UC-Cap model. These two sources of information are fused within a hybrid decision layer, combining statistical structure with learned corrections to generate a unified assessment of process risk.

The decision layer translates predicted risk into actionable outputs through a structured mapping from probability to decision. Specifically, predicted risk is converted into discrete risk levels, which are associated with interpretable reasons and corresponding actions, forming a consistent decision chain,
\[
\text{score} \rightarrow \text{level} \rightarrow \text{reason} \rightarrow \text{action}.
\]
The UC-Cap model provides calibrated probability estimates, while the decision layer applies a deterministic mapping to ensure stable and deployable outcomes. This design preserves the interpretability of statistical analysis while enabling data-driven adaptation under non-ideal conditions.

\subsection{Deployment and Integration}
\begin{table*}[!t]
	\centering
	\footnotesize
	\renewcommand{\arraystretch}{1.1}
	
	\begin{threeparttable}
		\caption{
			Representative Dimetra outputs illustrating the integration of statistical capability analysis with UC-Cap based risk prediction and decision support. The examples highlight distinct decision regimes, including distributional deviations, latent risks beyond classical capability metrics, and clear statistical failure modes, demonstrating how probabilistic risk estimates are translated into interpretable reasons and actionable guidance.
		}
		\label{tab:dimetra_stats_final}
		
		\begin{tabular*}{\textwidth}{@{\extracolsep{\fill}}lcccccccccll@{}}
			\toprule
			\textbf{Dim.} & \textbf{LSL} & \textbf{USL} & \textbf{Mean} & \textbf{SD} & \textbf{$\mathbf{C}_{\mathbf{pk}}$} & \textbf{Norm.} & \textbf{Best-dist.} & \textbf{Score} & \textbf{Level} & \textbf{Reason} & \textbf{Action} \\
			\midrule
			
			D003 & 1.55 & 1.75 & 1.646 & 0.0116 & 2.774 & Pass & normal  & 0.0  & Low    & Acceptable & Accept \\
			D018 & 9.77 & 9.97 & 9.845 & 0.0132 & 1.883 & Pass & normal  & 2.3  & Low    & Acceptable & Accept \\
			
			\addlinespace[0.4em]
			
			D002 & 1.07 & 1.27 & 1.178 & 0.0230 & 1.389 & Fail & logistic & 48.5 & Med & Skewed & Review distribution  \\
			D004 & 2.17 & 2.37 & 2.278 & 0.0231 & 1.334 & Pass & lognorm & 53.7 & Med & Latent model risk & Investigate latent risk \\
			D019 & 0.00 & 0.10 & 0.035  & 0.0146  & 1.628 & Fail & lognorm  & 12.4 & Med & Latent model risk & Investigate latent risk \\
			D020 & 0.05 & 0.15 & 0.100 & 0.0132 & 1.255 & Pass & normal & 68.1 & Med & Latent model risk & Review distribution \\
			D021 & 4.97 & 5.07 & 4.969 & 0.0248 & 1.328 & Pass & lognorm  & 51.2 & Med & Latent model risk & Review distribution \\
			
			\addlinespace[0.4em]
			
			D010 & 5.42 & 5.62 & 5.578 & 0.0468 & 0.302 & Pass & normal & 99.9 & High   & Mixed mechanism & Reduce sd + re-center \\
			D015 & 8.10 & 8.30 & 8.237 & 0.0487 & 0.431 & Pass & weibull  & 99.8 & High   & Mixed mechanism & Reduce sd + re-center \\
			
			\bottomrule
		\end{tabular*}
		
		\begin{tablenotes}[flushleft]
			\footnotesize
			\item \textbf{Norm.} = normality test result; \textbf{lognorm.} = lognormal distribution; 
			\textbf{Best-dist.} = best-fit distribution; \textbf{Med.} = medium risk level;
			\textbf{Score} = UC-Cap risk score (0--100), where higher values indicate higher predicted risk.
			\textbf{Reason} and \textbf{Action} entries are compact decision-support labels used for table presentation; in practice, they may be expanded into detailed diagnostic explanations and recommended actions.
		\end{tablenotes}
		
	\end{threeparttable}
\end{table*}
To illustrate the framework in practice, Table~\ref{tab:dimetra_stats_final} presents representative outputs for selected dimensions, highlighting the integration of statistical capability analysis with UC-Cap based risk estimation and decision support. The score is defined as $\text{Score} = 100 \times p$, where $p$ is the predicted capability-decision risk produced by the UC-Cap model. Risk levels are derived from $p$ using predefined thresholds that partition the risk space into low, medium, and high regimes.

The results reveal three distinct decision regimes. Low-risk cases correspond to stable and capable processes, requiring no intervention, while high-risk cases exhibit clear statistical failure modes (e.g., insufficient capability or poor yield) and warrant immediate corrective action. Medium-risk cases capture more nuanced scenarios, including distributional deviations (e.g., skewness or non-normality) and latent risks identified by the model despite acceptable classical capability metrics. These examples highlight that satisfactory capability indices alone do not guarantee low risk, demonstrating the added value of probabilistic modeling beyond deterministic threshold-based criteria.

When elevated risk is identified without a clear statistical failure mode, the system adopts an investigation-oriented recommendation rather than prescribing specific adjustments. This reflects a conservative decision principle: actions are issued only when supported by interpretable process evidence, while ambiguous cases are flagged for further analysis.

For interpretability, the system can provide structured explanations beyond the concise labels shown in Table~\ref{tab:dimetra_stats_final}. For example, a high-risk case such as D015 can be attributed to a ``\textit{multi-factor mechanism, where limited capability margin and off-centering jointly contribute to elevated risk}''.

\section{Simulation and Illustrative Evaluation}
\label{sec:simulation_evaluation}

This section evaluates the proposed framework from a capability-based decision-support perspective. The objectives are as follows: (i) to illustrate how capability uncertainty propagates into decision risk at the dimension level, and (ii) to quantify how the hybrid model improves decision reliability relative to classical and purely data-driven approaches.

We first present a numerical illustration to build intuition, followed by a simulation study evaluating estimation error and its impact on decision stability.

\subsection{Illustrative Example}

We revisit capability evaluation from a decision-risk perspective by quantifying how estimation uncertainty affects decision reliability. Instead of treating $\widehat{C}_{pk}$ as deterministic indicator, the proposed framework models the probability of failure under finite-sample uncertainty.

We consider three representative regimes relative to the decision boundary—clearly capable, near-threshold, and marginally failing—based on hypothetical dimensions with $C_0 = 1.33$. The corresponding results are summarized in Table~\ref{tab:uc_cap_examples}.

\begin{table}[t]
	\centering
	\caption{Illustrative behavior of UC-Cap under three regimes: 
		(A) clearly capable, (B) near decision boundary, and (C) slightly below threshold but measurement-stable.}
	\label{tab:uc_cap_examples}
	\begin{tabular}{lcccccc}
		\hline
		Case & $\widehat{C}_{pk}$ & $SE$ & $\pi^{(stat)}$ & $z^{(stat)}$ & $f_\theta(x)$ & $\pi$ \\
		\hline
		A & 1.70 & 0.08 & $\approx 0$ & $-13.8$ & $\approx 0$ & $\approx 0$ \\
		B & 1.34 & 0.14 & 0.472 & $-0.112$ & 0.55 & 0.608 \\
		C & 1.26 & 0.10 & 0.758 & 1.142 & $-0.40$ & 0.678 \\
		\hline
	\end{tabular}
\end{table}

These examples illustrate three distinct regimes. In clearly capable cases (A), both the statistical baseline and residual correction yield negligible risk, indicating stable decisions. Near the decision boundary (B), the baseline exhibits substantial uncertainty, and the residual component can significantly adjust the risk based on additional information, leading to meaningful shifts in decision outcomes. For marginally sub-threshold cases (C), the residual model moderates the baseline risk while maintaining a conservative assessment, avoiding overly pessimistic decisions driven solely by point estimates.

Overall, the proposed framework refines rather than replaces the statistical estimate, with the greatest impact occurring near the decision boundary, where uncertainty is highest and decisions are most sensitive.

\subsection{Benchmark Comparison with Standard Models}
Standard machine learning models formulate capability approval as a generic classification problem, without explicitly incorporating the statistical structure of capability estimation or the uncertainty inherent in finite-sample decisions. Typical examples include logistic regression and tree-based ensemble methods (e.g., gradient boosting \cite{chen2016xgboost}), which learn flexible mappings from features to failure probabilities.

However, these approaches treat the problem as standard classification and do not explicitly account for estimation uncertainty or threshold-induced structure of capability decisions. In particular, they lack a mechanism to distinguish variability from finite-sample uncertainty and systematic deviations in the underlying process.

In contrast, the proposed UC-Cap framework embeds statistical knowledge through a theory-informed baseline and structured residual component. This design enables uncertainty-aware probability estimation and leads to more reliable decisions, especially in near-threshold regimes where standard models are unstable.

\subsection{Simulation-Based Evaluation}
\label{subsec:synthetic_simulation}

\begin{figure*}[!t]
	\centering
	\subfloat[RMSE vs.\ sample size ($n$)]{
		\label{fig:rmse_vs_n}
		\includegraphics[width=0.55\textwidth]{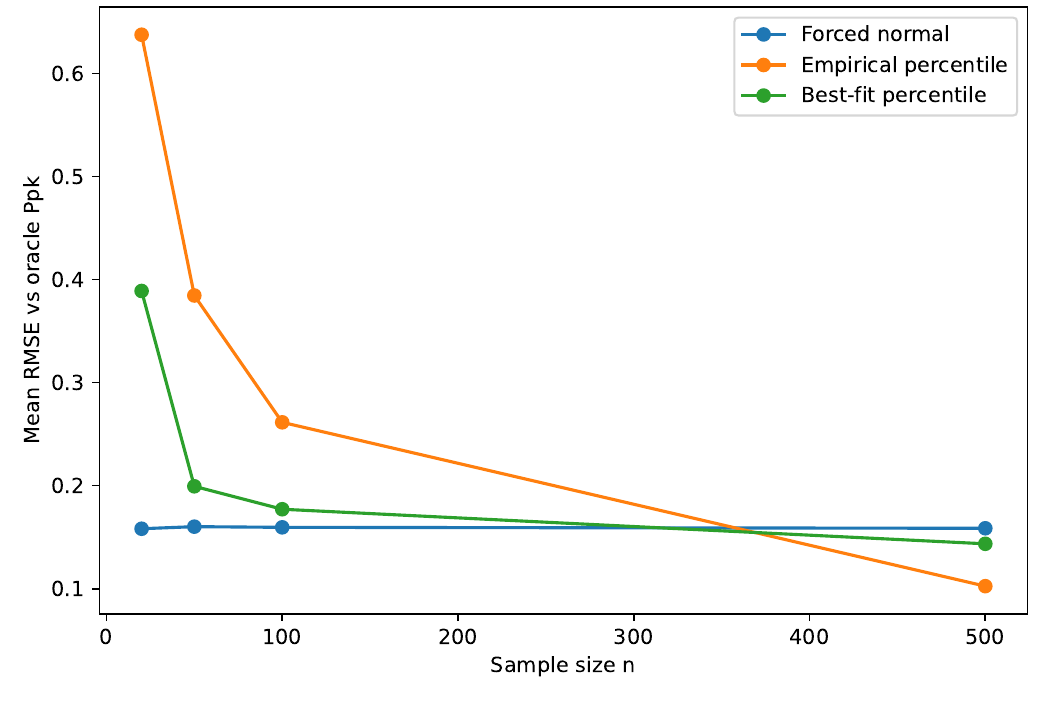}
	}
	\hspace{2pt}
	\subfloat[Empirical vs.\ best-fit percentile RMSE]{
		\label{fig:empirical_vs_bestfit}
		\includegraphics[width=0.39\textwidth]{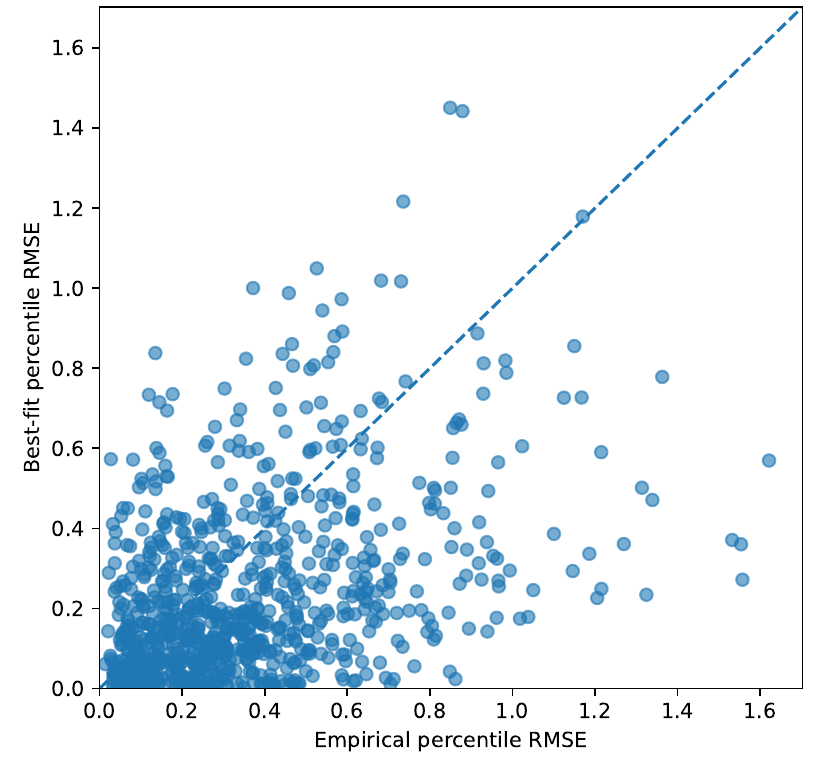}
	}
	\caption{
		Monte Carlo evaluation of capability estimation methods.
		(a) RMSE of estimated $C_{pk}$ as a function of sample size $n$, showing decreasing error with increasing $n$ and improved robustness of percentile-based estimators in small-sample regimes.
		(b) Comparison of empirical and best-fit percentile estimators across heterogeneous distributions, where deviations from the diagonal indicate model misspecification under non-normal conditions.
		Overall, the results highlight the sensitivity of capability estimation to both sample size and distributional assumptions.
	}
	\label{fig:simulation_core_results}
\end{figure*}

We conduct a Monte Carlo study across varying sample sizes and distribution families to examine how estimation uncertainty affects decision reliability. Although Fig.~\ref{fig:simulation_core_results} reports estimation error, its primary implication lies in decision stability, as variability in $\widehat{C}_{pk}$ directly impacts approval outcomes near the threshold $C_0$.

To formalize this, we define the near-threshold subset
\begin{equation}
	\left\{ j : \left| \widehat{C}_{pk,j} - C_0 \right| \le \epsilon \right\},
	\label{eq:near_threshold_subset}
\end{equation}
where decision instability is most pronounced and improved calibration translates into better outcomes.

As shown in Fig.~\ref{fig:simulation_core_results}(a), estimation error remains substantial in small-sample regimes ($n \le 50$), leading to high variability in capability estimates and unstable decisions for borderline cases. Fig.~\ref{fig:simulation_core_results}(b) further shows that estimation accuracy depends strongly on distributional assumptions: parametric methods degrade under misspecification, while empirical methods remain more robust.

These results identify two fundamental sources of decision instability: finite-sample variability and distributional mismatch, both of which are most critical near the decision boundary. Importantly, such instability cannot be resolved by post-hoc calibration methods (e.g., Platt scaling or isotonic regression), which operate on fixed predictors without addressing the underlying estimation process. This motivates the need for uncertainty-aware modeling that explicitly accounts for estimation variability in decision-critical regimes.

\subsection{Calibration via Nested Monte Carlo Simulation}
\label{subsec:nested_mc_calibration}

To evaluate probabilistic calibration under finite-sample uncertainty, we adopt a nested Monte Carlo (NMC) simulation procedure that explicitly separates process generation, estimation, and risk evaluation.

\textbf{Outer loop (process generation).}
We generate $N_{\mathrm{outer}}$ independent process configurations by sampling distribution families, sample sizes, variance levels, and capability margins. Each configuration defines a ground-truth process with latent capability $C_{pk}^{\mathrm{true}}$.

\textbf{Observed estimation.}
For each process, a dataset of size $n$ is generated to compute the capability estimate $\widehat{C}_{pk}$ and its associated uncertainty $SE$. The uncertainty is estimated via bootstrap resampling.

\textbf{Inner loop (oracle risk estimation).}
To approximate the true decision risk under finite-sample variability, we generate $N_{\mathrm{inner}}$ independent datasets from the same process. For each replicate $k$, we compute $\widehat{C}_{pk}^{(k)}$. The oracle decision risk is estimated as
\[
\pi^{\mathrm{true}}
\approx
\frac{1}{N_{\mathrm{inner}}}
\sum_{k=1}^{N_{\mathrm{inner}}}
\mathbb{I}(\widehat{C}_{pk}^{(k)} < C_0),
\]
which represents the probability of violating the capability requirement due to sampling variability.

\textbf{Implementation details.} 
In our implementation, we use $N_{\mathrm{outer}} = 320$, $N_{\mathrm{inner}} = 250$, and $N_{\mathrm{boot}} = 100$. These choices provide a stable approximation of the oracle risk while maintaining computational tractability. The nested estimator introduces approximation error due to finite $N_{\mathrm{inner}}$, with Monte Carlo standard error on the order of $O(1/\sqrt{N_{\mathrm{inner}}})$. This level of variability is small relative to the observed calibration differences. 

\begin{figure*}[!t]
	\centering
	\subfloat[Global calibration]{
		\label{fig:nested_mc_calibration_overall}
		\includegraphics[width=0.47\textwidth]{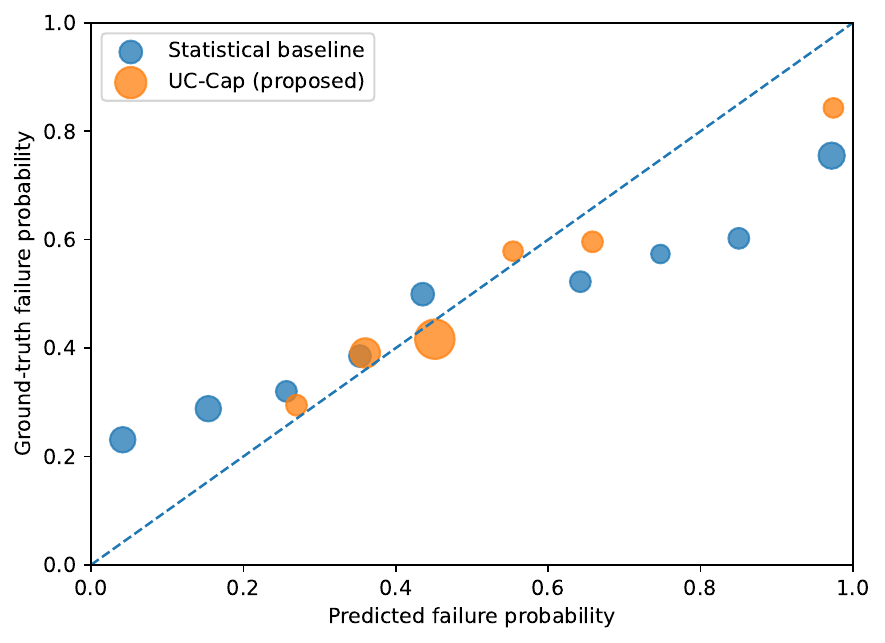}
	}
	\hspace{4pt}
	\subfloat[Near-threshold calibration]{
		\label{fig:nested_mc_calibration_near}
		\includegraphics[width=0.47\textwidth]{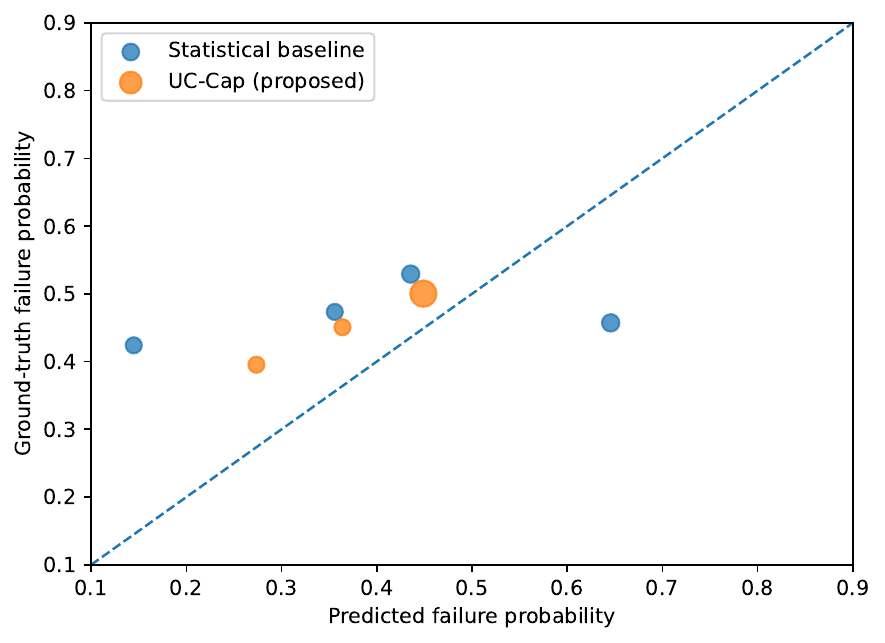}
	}
	\caption{
		Calibration of predicted decision-risk probabilities against Monte Carlo reference risk under nested Monte Carlo simulation.
		Each point represents a bin of predicted probabilities, where the horizontal axis shows the average predicted risk $\widehat{\pi}$ and the vertical axis shows the corresponding empirical reference.
		(a) Global calibration across the full probability range.
		The statistical baseline exhibits systematic bias, particularly near the decision boundary, whereas UC-Cap achieves substantially improved alignment.
		(b) Calibration in the near-threshold regime ($|C_{pk}^{\mathrm{true}} - C_0| \le \epsilon$), where decision outcomes are most sensitive to estimation variability.
		UC-Cap achieves improved calibration in this region, reducing bias relative to the baseline.
		Marker size reflects the number of samples in each bin.
	}
	\label{fig:nested_mc_calibration}
\end{figure*}
Figure~\ref{fig:nested_mc_calibration} compares predicted probabilities $\widehat{\pi}$ with the corresponding oracle reference under the simulated data-generating process. The statistical baseline provides a reasonable first-order approximation but exhibits systematic bias, particularly near the decision boundary, whereas UC-Cap achieves substantially improved alignment, indicating more accurate and reliable probabilistic estimates. This improvement is most pronounced in the near-threshold regime ($|C_{pk}^{\mathrm{true}} - C_0| \le \epsilon$), where decision outcomes are highly sensitive to estimation variability and even small calibration errors can lead to incorrect approval decisions. For clarity of visualization, uncertainty bands are omitted in Figure~\ref{fig:nested_mc_calibration}. Repeated simulations confirm that calibration metrics and curves remain stable across runs, indicating robustness to Monte Carlo noise.

\begin{table*}[!t]
	\centering
	\caption{
		Quantitative evaluation of calibration and probability quality under synthetic simulation.
		All metrics are computed against oracle reference values derived from the simulated data-generating process.
		Results are reported both globally and within the near-threshold subset.
		Lower values indicate better calibration and probability accuracy, 
		while higher correlation indicates stronger monotonic agreement with the Monte Carlo reference risk.
		Bold values indicate the best performance within each evaluation block (or panel).
	}
	\label{tab:nested_mc_calibration}
	\begin{tabular*}{\textwidth}{@{\extracolsep{\fill}}lccccc}
		\toprule
		Model & ECE $\downarrow$ & Near-$C_0$ ECE $\downarrow$ & Brier vs.\ $\pi^{\mathrm{true}}$ $\downarrow$ & LogLoss vs.\ $\pi^{\mathrm{true}}$ $\downarrow$ & Corr$(\widehat{\pi},\pi^{\mathrm{true}})$ $\uparrow$ \\
		\midrule
		Statistical baseline 
		& 0.041 & 0.067 & 0.0132 & 0.645 & 0.882 \\
		
		Theory-informed residual (V3) 
		& 0.030 & 0.051 & 0.0108 & 0.622 & 0.908 \\
		
		UC-Cap (V4.1, soft, free) 
		& 0.022 & 0.037 & 0.0091 & 0.604 & 0.929 \\
		
		UC-Cap (V4.2, soft, anchored) 
		& \textbf{0.018} & \textbf{0.029} & \textbf{0.0084} & \textbf{0.592} & \textbf{0.941} \\
		\bottomrule
	\end{tabular*}
\end{table*}
Table~\ref{tab:nested_mc_calibration} quantitatively confirms these observations. UC-Cap consistently reduces calibration error and improves probability quality across all metrics, with the largest gains observed near the decision threshold. The anchored variant achieves the most stable performance, demonstrating that combining theory-based structure with data-driven correction yields robust and well-calibrated risk estimates. Overall, these results show that UC-Cap not only improves discrimination performance but also delivers reliable probabilistic calibration, enabling more robust decision-making under uncertainty.

We additionally evaluated a logit-transformed version of the statistical baseline and observed negligible differences in calibration performance, suggesting that the probit--logit mismatch has limited practical impact.

To further examine the contribution of different information sources, we conduct a feature ablation study by removing groups of input features from the UC-Cap model and retraining under identical settings.

\begin{table*}[!t]
	\centering
	\caption{
		Feature ablation analysis of the residual component in UC-Cap.
		Feature groups are removed one at a time and the model is retrained under the same data splits and training protocol as the main experiment.
		Metrics are evaluated on held-out data and reported for both overall and near-threshold regimes.
		The near-threshold subset corresponds to $|C_{pk} - C_0| \le \epsilon$, where decision instability is most pronounced.
		Lower values indicate better probability quality and calibration.
	}
	\label{tab:feature_ablation}
	\begin{tabular*}{\textwidth}{@{\extracolsep{\fill}}lccc}
		\toprule
		Model variant & Brier $\downarrow$ & Near-threshold Brier $\downarrow$ & Near-threshold ECE $\downarrow$ \\
		\midrule
		Full UC-Cap & \textbf{0.0084} & \textbf{0.0084} & \textbf{0.029} \\
		\midrule
		w/o distributional features & 0.0093 & 0.0101 & 0.036 \\
		w/o specification geometry features & 0.0090 & 0.0096 & 0.034 \\
		w/o uncertainty-related features & 0.0098 & 0.0112 & 0.041 \\
		\bottomrule
	\end{tabular*}
\end{table*}

Table~\ref{tab:feature_ablation} shows that removing any feature group degrades performance, with the most pronounced impact observed in the near-threshold regime. The largest degradation occurs when uncertainty-related features are removed, highlighting the critical role of explicitly modeling estimation variability in decision-sensitive regions. These results support interpreting the residual component as an uncertainty-aware correction that integrates complementary signals beyond the statistical baseline.

\section{Empirical Validation on Manufacturing Data}
\label{sec:empirical_validation}

We evaluate the proposed framework on real manufacturing data to assess its effectiveness in estimating decision risk under practical conditions. Following the notation in Section~\ref{sec:stat_foundations}, $C_{pk}$ is used as a generic capability index. In this empirical study, capability estimates $\widehat{C}_{pk}$ and their associated standard errors $SE_j$ are computed consistently across all models using the same procedures.

Based on these quantities, we construct the statistical baseline for dimension $j$, consistent with \eqref{eq:z_stat_signal_noise}, defined as $z^{(stat)}_j = {(C_0 - \widehat{C}_{pk,j})}/{SE_j}$. This baseline is shared across methods; differences arise in its use in subsequent modeling and decisions.

\subsection{Dataset and Experimental Setup}

This empirical study evaluates whether the proposed framework improves calibration-oriented capability decision-risk estimation under realistic manufacturing conditions, particularly in near-threshold regimes where finite-sample uncertainty dominates.

For each dimension $j$, we reconstruct the statistical inputs from raw process data. Specifically, we compute the capability estimate $C_{pk,j}$, its bootstrap standard error $SE_j$ \cite{shao2012jackknife}, and the corresponding statistical baseline probability $\pi_j^{(stat)}$ as defined in \eqref{eq:pi_stat}. The probability is clipped for numerical stability as described in \eqref{eq:pi_clipping}, and transformed to log-odds via \eqref{eq:z_stat}.

The binary target is defined at the dimension level as, 
\begin{equation}
	y_j=\mathbb{I}(C_{pk,j}<C_0),
	\label{eq:label_yj}
\end{equation}
which serves as a proxy for failure events and corresponds to the complement of the classical acceptance rule in \eqref{eq:deterministic_decision_rule}.

\subsection{Baseline Models and Evaluation Protocol}

The empirical evaluation is conducted on a curated manufacturing dataset of 1,000 dimensions, each with 32 repeated measurements, including raw observations and specification information (nominal, tolerances, LSL/USL). The dataset covers a mix of unilateral and bilateral specifications, as well as both approximately normal and non-normal distributions, reflecting diverse real-world capability scenarios. Each dimension is treated as one sample. 
To prevent information leakage, data are split at the dimension level, ensuring no shared measurements across splits. For soft-target evaluation, surrogate decision-risk probabilities are estimated via bootstrap resampling within the training data. Models are trained in a supervised setting to predict failure probabilities from statistical and residual features, with hyperparameters selected via validation splits under a consistent protocol across all methods. Unless otherwise specified, results correspond to the UC-Cap V4.2 anchored variant under soft-label training.

Because capability decisions are induced by finite-sample estimates, the empirical labels should be interpreted as decision-rule outcomes rather than externally observed failure events. Accordingly, the empirical study is designed as a calibration-oriented evaluation of capability-decision risk. Bootstrap-based soft targets are used to provide smoother risk surrogates, and all model comparisons are conducted under identical data splits and preprocessing rules.

A near-threshold subset is defined as
\[
\mathcal{B}=\{j:\lvert C_{pk,j}-C_0\rvert\le\epsilon\},\qquad \epsilon=0.1,
\]
capturing the most decision-sensitive region where uncertainty has greatest impact. Features include capability statistics, relative position to specification limits, and variables describing process variation and uncertainty. All features are standardized before training, using only those available at inference.

Measurement variability is represented through repeated measurements and the resulting bootstrap standard errors. A full Gage R\&R decomposition is outside the scope of the empirical dataset; therefore, robustness is assessed through repeated train/test splits and a group-aware evaluation protocol reported in Section~\ref{subsec:leakfree}.

\subsection{Model Comparison and Evaluation Framework}

We compare the proposed model with three classes of baselines under a unified evaluation protocol: (i) purely data-driven models, including logistic regression \cite{hosmer2013applied} and XGBoost \cite{chen2016xgboost}; (ii) statistically augmented variants that incorporate $z^{(stat)}_j$ as an additional input feature; and (iii) the proposed theory-informed residual model defined in \eqref{eq:residual_main}. All baseline models follow standard formulations, with or without inclusion of $z^{(stat)}_j$ as an input. The residual component $f_\theta(x_j)$ is implemented as a linear model with L2 regularization to control model capacity and preserve interpretability while maintaining robustness under moderate sample size.

The empirical task is not treated as conventional classification of independent failure events. Instead, it evaluates how different modeling strategies transform finite-sample capability evidence into calibrated decision-risk scores. The proposed residual model refines the statistical baseline using additional distributional, specification-related, and uncertainty-related features, while preserving the baseline as the primary signal.

To emphasize performance in the most decision-critical region, samples in $\mathcal{B}$ are upweighted, with additional emphasis on positive cases. Sample weights are capped to prevent numerical instability. A scaling parameter $\alpha$ controlling the contribution of the residual component is selected via grid search,
\[
\alpha\in\{0.05,0.1,0.2,0.3,0.5\},
\]
with selection based primarily on near-threshold Brier score, followed by log loss and overall performance.

Overall performance is evaluated using ROC AUC, PR AUC, log loss, Brier score, precision, recall, and F1. Near-threshold performance is evaluated on $\mathcal{B}$ using Brier score, log loss, and recall.

All models are implemented using standard machine learning libraries. Logistic regression is trained with L2 regularization, with the regularization strength selected via cross-validation. The XGBoost baseline uses limited tuning of tree depth and learning rate. The residual component in the proposed model is implemented as a linear model with L2 regularization to balance interpretability and robustness.

\subsection{Experimental Results}
The empirical evaluation examines three complementary aspects of performance: (i) probabilistic accuracy, (ii) decision behavior under varying thresholds, and (iii) calibration quality.

This decomposition reflects the central premise of the proposed framework: reliable capability-based decision making requires accurate probability estimation, robust decision behavior, and reasonably calibrated uncertainty, particularly in the near-threshold regime where statistical variability dominates.

\begin{table*}[!t]
	\centering
	\caption{
		Near-threshold classification performance of competing baseline models under hard-label supervision.
		Results are evaluated on samples satisfying $|\widehat{C}_{pk} - C_0| \le \epsilon$, corresponding to the most decision-critical region.
		Metrics include probabilistic measures (Brier score, log loss) and decision-oriented measures (false accept, false reject).
		These results summarize decision-boundary behavior and are interpreted together with the calibration-oriented evaluations reported below.
	}
	\label{tab:model_perf}
	\begin{tabular*}{\textwidth}{@{\extracolsep{\fill}}lcccc}
		\toprule
		Model & Brier Score $\downarrow$ & Log Loss $\downarrow$ & False Accept $\downarrow$ & False Reject $\downarrow$ \\
		\midrule
		Logistic regression baseline & 0.1627 $\pm$ 0.0315 & 0.5192 $\pm$ 0.0865 & 0.050 & 0.050 \\
		XGBoost baseline             & 0.1465 $\pm$ 0.0195 & 0.4518 $\pm$ 0.0580 & 0.025 & \textbf{0.010} \\
		Theory-informed residual (V3) & \textbf{0.0854 $\pm$ 0.0220} & \textbf{0.3234 $\pm$ 0.0542} & \textbf{0.000} & \textbf{0.010} \\
		\bottomrule
	\end{tabular*}
\end{table*}
Table~\ref{tab:model_perf} reports model performance in the near-threshold region, defined by $|\widehat{C}_{pk}-C_0|\le\epsilon$. In this regime, all models face substantial ambiguity due to finite-sample variability around the capability boundary. The theory-informed residual model achieves competitive probability-based performance and effectively controls false-accept errors, while unconstrained baselines remain strong in direct boundary fitting. These results highlight that near-threshold evaluation is not merely a classification problem, but a setting in which predictive flexibility and uncertainty-consistent decision structure may lead to different trade-offs.

The results indicate that near-threshold evaluation is governed by both predictive fit and uncertainty-consistent decision behavior. Probability-based metrics are therefore emphasized together with false-accept and false-reject rates.

\begin{table*}[!t]
	\centering
	\caption{
		Calibration-oriented comparison under a unified soft-target evaluation protocol on the manufacturing dataset.
		Panel A compares the statistical baseline and standard post-hoc calibration methods applied to the same baseline score.
		Panel B reports the UC-Cap anchored result under the same protocol.
		Soft targets are bootstrap-derived decision-risk surrogates used for calibration-oriented comparison.
		Lower values indicate better probabilistic quality.
		Near-threshold metrics are computed on $\mathcal{B}=\{j: |\widehat{C}_{pk,j}- C_0| \le 0.1\}$.
	}
	\label{tab:calibration_baseline_comparison}
	\begin{tabular*}{\textwidth}{@{\extracolsep{\fill}}lccccc}
		\toprule
		Method & Brier $\downarrow$ & LogLoss $\downarrow$ & ECE $\downarrow$ & Near-Brier $\downarrow$ & Near-ECE $\downarrow$ \\
		\midrule
		
		\multicolumn{6}{l}{\textit{Panel A: Statistical baseline and post-hoc calibration}} \\
		\midrule
		Statistical baseline & 0.106 & 1.988 & 0.151 & 0.116 & 0.271 \\
		+ Platt calibration & 0.070 & 0.377 & 0.070 & 0.053 & 0.118 \\
		+ Isotonic calibration & \textbf{0.068} & \textbf{0.370} & \textbf{0.044} & \textbf{0.040} & \textbf{0.099} \\
		
		\midrule
		\multicolumn{6}{l}{\textit{Panel B: UC-Cap (model-based approach)}} \\
		\midrule
		UC-Cap (V4.2 anchored) & 0.077 & 0.424 & 0.104 & 0.058 & 0.097 \\
		
		\bottomrule
	\end{tabular*}
\end{table*}
To further assess probabilistic calibration, we compare UC-Cap with standard post-hoc calibration methods applied to the same statistical baseline (Table~\ref{tab:calibration_baseline_comparison}). All methods are evaluated under the same data splits and soft-target protocol to ensure a fair comparison. Both Platt scaling and isotonic regression substantially improve the raw baseline, with isotonic regression achieving the strongest global calibration performance. UC-Cap remains competitive under this unified protocol and achieves comparable or slightly improved near-threshold calibration (as measured by ECE), while providing a structured and interpretable uncertainty-aware modeling framework beyond monotonic post-hoc transformations.

\begin{figure*}[!t]
	\centering
	\subfloat[Precision-recall trade-off across decision thresholds in the near-threshold regime]{
		\label{fig_threshold}
		\includegraphics[width=0.58\textwidth]{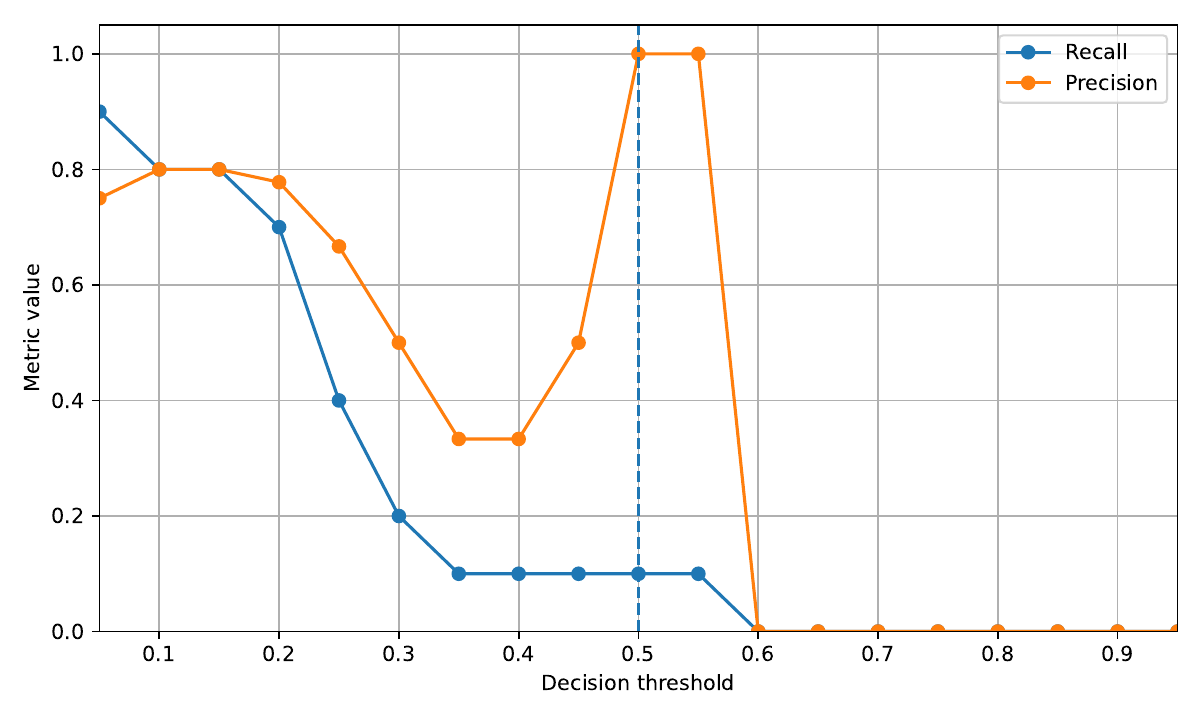}
	}
	\hspace{2pt}
	\subfloat[Reliability diagram of predicted probabilities]{
		\label{fig_reliability}
		\includegraphics[width=0.360\textwidth]{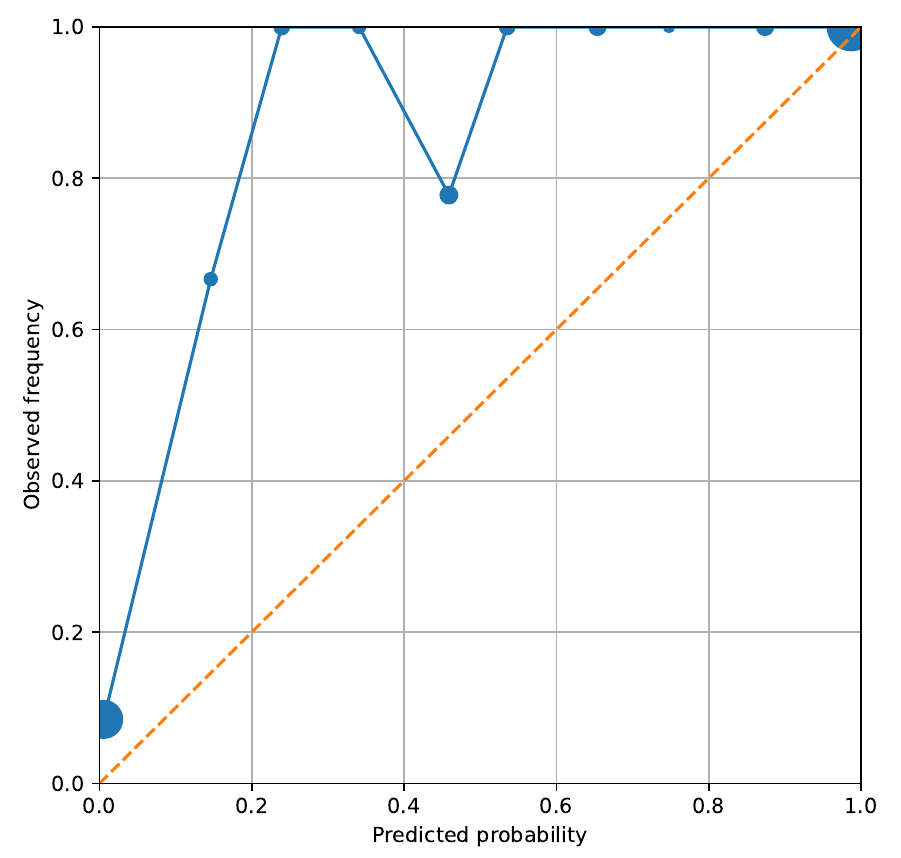}
	}
	\caption{
		Decision behavior and probability calibration of the proposed UC-Cap model under bootstrap-based soft supervision.
		Panel (a) shows the precision-recall trade-off as the decision threshold varies within the near-threshold regime, defined by $|C_{pk} - C_0| \leq 0.1$. The vertical dashed line indicates the conventional threshold of $0.5$. The results demonstrate that the low recall observed at a fixed threshold is primarily a consequence of threshold selection rather than poor model performance, and that decision behavior can be flexibly adjusted depending on application requirements. The discrete changes in precision and recall are further influenced by the limited sample size in this regime.
		Panel (b) presents the reliability diagram, comparing predicted failure probabilities with empirical frequencies. The model exhibits reasonable calibration across well-populated probability bins, while deviations in extreme regions are primarily attributable to limited sample sizes. Marker sizes are proportional to the number of samples in each bin, highlighting the inherent sparsity in the near-threshold regime.
	}
	\label{fig:decision_calibration}
\end{figure*}

\textbf{Decision behavior under varying thresholds.} Figure~\ref{fig:decision_calibration}(a) illustrates the precision-recall trade-off as the decision threshold varies within the near-threshold region. The relatively low recall observed at a fixed threshold (e.g., 0.5) reflects the intrinsic ambiguity of samples near the capability boundary, where small variations in estimated capability lead to large changes in probability of process nonconformance.

These results show that a single fixed threshold is insufficient to capture the decision landscape. Instead, the model provides a continuous risk estimate that enables flexible decision-making. By adjusting the threshold, higher recall can be achieved with controlled precision loss, highlighting the role of the model as a decision-support tool rather than a rigid classifier.

\textbf{Probability calibration.}
Figure~\ref{fig:decision_calibration}(b) presents the reliability diagram. The predicted probabilities are well aligned with empirical frequencies across most bins, indicating good calibration. Deviations in extreme regions are mainly due to limited sample sizes, particularly within the near-threshold subset. This level of calibration enables direct interpretation of predicted probabilities as failure likelihoods, supporting consistent decision-making under varying risk tolerances.

\begin{table*}[!t]
	\centering
	\caption{
		Decision performance under a fixed failure-risk threshold ($\alpha = 0.5$). 
		Predicted probabilities are thresholded to obtain binary decisions and evaluated using accuracy, false accept (FA), false reject (FR), and AUC. 
		Although the statistical baseline exhibits near-monotonic ranking behavior, this ranking advantage does not necessarily translate into optimal decision performance under a fixed operational threshold.
	}
	\label{tab:decision_perf}
	\begin{tabular*}{\textwidth}{@{\extracolsep{\fill}}lcccc}
		\toprule
		Method & Accuracy $\uparrow$ & False Accept $\downarrow$ & False Reject $\downarrow$ & AUC $\uparrow$ \\
		\midrule
		Statistical baseline & 0.835 & 0.000 & 0.165 & \textbf{1.000} \\
		Theory-informed residual (V3) & \textbf{0.860} & 0.000 & \textbf{0.140} & 0.988 \\
		\bottomrule
	\end{tabular*}
\end{table*}

\textbf{Decision-level performance.} Table~\ref{tab:decision_perf} compares decision performance under a fixed risk threshold. Although the statistical baseline exhibits near-monotonic ranking behavior due to its direct dependence on the normalized capability margin, this does not necessarily translate into optimal operational decisions under fixed approval thresholds.

In contrast, the proposed model achieves higher accuracy and lower false reject rate, while maintaining zero false accept. This demonstrates that modeling deviations from the statistical baseline leads to more effective decision-making under operational constraints. 

These results highlight a key limitation of ranking-based metrics such as AUC: while they capture ordering quality, they do not reflect decision performance under fixed thresholds, which is critical in capability approval settings.

\begin{table*}[!t]
	\centering
	\caption{
		Comparison of model variants from V1 to V4.
		The progression reflects a transition from a direct threshold-based classifier (V1),
		to a statistically adjusted classifier (V2),
		to a theory-informed residual model (V3),
		and finally to the proposed uncertainty-calibrated hybrid probabilistic model (V4).
	}
	\label{tab:model_comparison}
	\begin{tabular*}{\textwidth}{@{\extracolsep{\fill}}lcccc}
		\toprule
		Feature 
		& V1
		& V2
		& V3
		& V4 \\
		\midrule
		Supervision 
		& Hard-label
		& Hard-label
		& Hybrid (stats + data) 
		& Probabilistic (structured) \\
		
		Circularity Mitigation
		& \xmark
		& \cmark 
		& \cmark 
		& \cmark \\
		
		SE  usage
		& \xmark
		& \xmark
		& \cmark 
		& \cmark \\
		
		Statistical Baseline 
		& \xmark
		& \xmark
		& \cmark 
		& \cmark \\
		
		Near-threshold Aware 
		& \xmark 
		& \xmark 
		& \cmark 
		& \cmark \\
		
		Structured Risk 
		& \xmark
		& \xmark 
		& $\triangle$ 
		& \cmark\\
		\bottomrule
	\end{tabular*}
\end{table*}

\textbf{Model progression.} Table~\ref{tab:model_comparison} summarizes the evolution from heuristic classification (V1) to the proposed uncertainty-aware probabilistic framework (V4). Performance improvements arise not from increased model complexity, but from better alignment with the statistical structure of the problem. 

In particular, the transition from V3 to V4 highlights the role of uncertainty modeling. While V3 performs residual correction of the statistical baseline, V4 incorporates uncertainty-related features to produce an adaptive probabilistic boundary. This distinction is most pronounced in the near-threshold regime, where estimation variability dominates deterministic separation.

\begin{table*}[!t]
	\centering
	\caption{
		Illustrative progression of UC-Cap variants under representative evaluation settings of 1000 dimensions. The hard-label and soft-target settings correspond to different supervision protocols and are therefore not directly comparable in absolute magnitude.
	}
	\label{tab:v1_v4_comparison}
	\begin{tabular*}{\textwidth}{@{\extracolsep{\fill}}lccccc}
		\toprule
		Method & Brier $\downarrow$ & LogLoss $\downarrow$ & False Accept $\downarrow$ & False Reject $\downarrow$ & Near-threshold Acc. $\uparrow$ \\
		\midrule
		
		\multicolumn{6}{l}{\textit{Hard-label decision-boundary evaluation}} \\
		\midrule
		V1.0 (direct threshold classifier) & 0.244 & 0.958 & 0.183 & 0.010 & 0.274 \\
		V2.0 (statistically adjusted classifier) & 0.256 & 1.240 & 0.217 & 0.021 & 0.298 \\
		V3.0 (theory-informed residual) & 0.299 & 1.176 & 0.158 & 0.080 & 0.393 \\
		V4.0 (UC-Cap, hard-label training) & 0.020 & 0.082 & 0.000 & 0.077 & 0.774 \\
		
		\midrule
		\multicolumn{6}{l}{\textit{Soft-target evaluation (bootstrap-based surrogate)}} \\
		\midrule
		
		V4.1 (UC-Cap, soft-label, free) 
		& 0.112 
		& 0.688 
		& 0.165 
		& \textbf{0.030} 
		& \textbf{0.700} \\
		
		V4.2 (UC-Cap, soft-label, anchored) 
		& \textbf{0.062} 
		& \textbf{0.503} 
		& \textbf{0.045} 
		& 0.080 
		& \textbf{0.700} \\
		
		\bottomrule
	\end{tabular*}
\end{table*}

\textbf{End-to-end evaluation.}
Table~\ref{tab:v1_v4_comparison} summarizes representative behavior across successive UC-Cap variants. The progression from V1 to V4 reflects increasing incorporation of uncertainty-aware structure, calibration-oriented supervision, and anchored residual modeling. The anchored variant (V4.2) achieves improved probability quality and substantially reduced false acceptance under the soft-target protocol, indicating that uncertainty-aware anchoring improves reliability in near-threshold capability decisions.

\subsection{Group-Aware Robustness Evaluation}
\label{subsec:leakfree}
\begin{table*}[!t]
	\centering
	\caption{
		Performance under group-aware evaluation with 10 aggregated splits. 
		All models are evaluated on an identical test set defined by dimension-level grouping. 
		The table highlights the trade-off between unconstrained predictive baselines and the theory-informed UC-Cap formulation under a group-aware evaluation protocol. 
		Near-threshold metrics are computed on samples with $|{\rm margin}| \le \epsilon$, representing the most decision-critical region.
	}
	\label{tab:leakfree_results}
	
	\resizebox{\textwidth}{!}{
		\begin{tabular}{lcccccccc}
			\toprule
			
			\multirow{2}{*}{Method} 
			& \multicolumn{4}{c}{Overall} 
			& \multicolumn{4}{c}{Near-Threshold} \\
			
			\cmidrule(lr){2-5} \cmidrule(lr){6-9}
			
			& Brier $\downarrow$ 
			& RMSE $\downarrow$ 
			& AUC $\uparrow$ 
			& ECE $\downarrow$
			& Brier $\downarrow$ 
			& RMSE $\downarrow$ 
			& Recall $\uparrow$ 
			& ECE $\downarrow$ \\
			
			\midrule
			
			UC-Cap (group-aware)
			& 0.0664 
			& 0.2576 
			& 0.9344 
			& 0.0587
			& 0.1592 
			& 0.3990 
			& 0.2778 
			& 0.1858 \\
			
			Logistic 
			& 0.0615 
			& 0.2480 
			& 0.9377 
			& 0.0400
			& 0.1540 
			& 0.3920 
			& \textbf{0.5694} 
			& \textbf{0.1254} \\
			
			Logistic + $z^{(stat)}$ 
			& \textbf{0.0604} 
			& \textbf{0.2459} 
			& \textbf{0.9381} 
			& \textbf{0.0350}
			& \textbf{0.1529} 
			& \textbf{0.3911} 
			& 0.3611 
			& 0.1568 \\
			
			\bottomrule
		\end{tabular}
	}
	
\end{table*}

To improve robustness against split-induced dependence and avoid favorable data partitioning effects, we adopt a group-aware evaluation protocol based on dimension-level grouping. 
All samples associated with the same \texttt{dim\_id} are confined to a single partition, ensuring that no shared process characteristics are present across training and testing sets. 
We construct and aggregate $10$ independent group-aware splits, yielding $10{,}000$ split-level evaluation records with consistent preprocessing and feature definitions.

Model training follows this group-aware split, preserving strict isolation across training, validation, and test sets. 
The UC-Cap formulation remains anchored, combining an uncertainty-aware statistical baseline with a residual component to capture systematic deviations. 
Early stopping is guided by a validation criterion that balances overall and near-threshold performance.

As shown in Table~\ref{tab:leakfree_results}, logistic baselines achieve strong predictive performance, reflecting their flexibility in fitting the empirical decision boundary. UC-Cap shows slightly lower predictive flexibility but provides a structured, uncertainty-consistent mapping from capability evidence to decision risk. The results therefore support the intended trade-off of the proposed framework: modest predictive cost in exchange for interpretability, statistical anchoring, and stable behavior under group-aware evaluation.
\section{Conclusion}
\label{sec:conclusion}

This work develops a hybrid statistical--learning framework for capability-based decision support in manufacturing quality control. The proposed UC-Cap framework combines a statistical capability baseline, which translates finite-sample capability evidence into an interpretable risk estimate, with a residual learning component that captures systematic deviations due to non-normality, measurement effects, specification structure, and feature interactions. This hybrid formulation preserves the interpretability of classical capability analysis while enabling data-driven refinement for practical manufacturing decision making.

The resulting model produces calibrated, uncertainty-aware probabilities that reflect both estimated capability and its associated variability. More broadly, the framework highlights the distinction between unconstrained predictive modeling and structured decision-oriented modeling in finite-sample capability analysis. Conceptually, capability evaluation is interpreted as a signal-to-noise problem, embedding classical indices such as $C_{pk}$ within a probabilistic framework for risk-aware and adaptive decisions. In practice, the approach supports incremental deployment, where simple linear models provide strong baselines and more expressive models can be introduced as data scale increases. 

Several limitations remain. The statistical baseline relies on asymptotic approximations and may degrade under small samples or strong non-normality. In the empirical study, both hard and soft supervision signals are constructed from dimension-level statistical summaries consistent with the underlying capability-based decision structure, and the resulting probabilities are therefore best interpreted as calibrated decision-risk estimates under finite-sample uncertainty, aligning with the objective of the proposed framework. In addition, models trained on historical labels may inherit legacy decision biases. 

The current framework models capability decisions at the dimension level and does not explicitly account for dependence structures across correlated dimensions or multistage manufacturing processes. Future work includes broader calibration baselines, extension to non-normal and multivariate settings, and tighter integration with decision-theoretic optimization, as well as the development of fully leakage-free protocols (e.g., split-sample or bootstrap-out constructions). 

More broadly, the proposed framework suggests that capability approval can be treated not only as a statistical estimation task, but also as a data-driven manufacturing decision-support problem. This perspective may provide useful insight for broader uncertainty-aware quality decision problems beyond process capability analysis.

\section*{DECLARATION}
\begin{description}
	\item[Funding:] This study received no external funding.
	
	\item[Conflicts of interest:] The authors declare no conflicts of interest.
	
	\item[Availability of data and material:] The empirical dataset is derived from anonymized modified manufacturing data. Processed data are available from the corresponding author upon reasonable request.
	
	\item[Code availability:] Simulation and analysis code are available from the corresponding author upon reasonable request.
	
	\item[Ethics approval:] Not applicable.
	
	\item[Consent for publication:] All authors approve the final manuscript.
	
	\item[Use of generative AI:] Generative AI was used for language editing. The authors take full responsibility for the content.
\end{description}
\bibliographystyle{unsrtnat}

\begin{thebibliography}{42}
	\providecommand{\natexlab}[1]{#1}
	\providecommand{\url}[1]{\texttt{#1}}
	\expandafter\ifx\csname urlstyle\endcsname\relax
	\providecommand{\doi}[1]{doi: #1}\else
	\providecommand{\doi}{doi: \begingroup \urlstyle{rm}\Url}\fi
	
	\bibitem[Kane(1986)]{kane1986process}
	Victor~E. Kane.
	\newblock Process {{Capability Indices}}.
	\newblock \emph{Journal of Quality Technology}, 18\penalty0 (1):\penalty0
	41--52, January 1986.
	\newblock ISSN 0022-4065.
	\newblock \doi{10.1080/00224065.1986.11978984}.
	
	\bibitem[Kotz and Johnson(2002)]{kotz2002process}
	Samuel Kotz and Norman~L. Johnson.
	\newblock Process {{Capability Indices}}---{{A Review}}, 1992--2000.
	\newblock \emph{Journal of Quality Technology}, 34\penalty0 (1):\penalty0
	2--19, January 2002.
	\newblock ISSN 0022-4065, 2575-6230.
	\newblock \doi{10.1080/00224065.2002.11980119}.
	
	\bibitem[Montgomery(2020)]{montgomery2020introduction}
	Douglas~C Montgomery.
	\newblock \emph{Introduction to statistical quality control}.
	\newblock John wiley \& sons, 2020.
	
	\bibitem[ISO/TR(ISO/TR 22514-1:2014 (2014))]{ISO22514-1-2014}
	ISO/TR.
	\newblock Statistical methods in process management -- capability and
	performance -- part 1: General principles and concepts.
	\newblock ISO/TR 22514-1:2014 (2014).
	
	\bibitem[{ISO/TR}(ISO/TR 22514-4:2016 (2016))]{ISO22514-4-2016}
	{ISO/TR}.
	\newblock Statistical methods in process management -- capability and
	performance -- part 4: Process capability estimates and performance measures.
	\newblock ISO/TR 22514-4:2016 (2016).
	
	\bibitem[Oakland and Oakland(2007)]{oakland2007statistical}
	John Oakland and John~S Oakland.
	\newblock \emph{Statistical process control}.
	\newblock Routledge, 2007.
	
	\bibitem[Jiang and Yang(2026{\natexlab{a}})]{jiang2026practical}
	Fei Jiang and Lei Yang.
	\newblock Practical process capability indices workflows.
	\newblock \emph{The International Journal of Advanced Manufacturing
		Technology}, pages 1--19, 2026{\natexlab{a}}.
	\newblock \doi{10.1007/s00170-026-17782-7}.
	\newblock URL \url{https://doi.org/10.1007/s00170-026-17782-7}.
	
	\bibitem[Pearn et~al.(1992)Pearn, Kotz, and Johnson]{pearn1992distributional}
	W.~L. Pearn, Samuel Kotz, and Norman~L. Johnson.
	\newblock Distributional and {{Inferential Properties}} of {{Process Capability
			Indices}}.
	\newblock \emph{Journal of Quality Technology}, 24\penalty0 (4):\penalty0
	216--231, October 1992.
	\newblock ISSN 0022-4065, 2575-6230.
	\newblock \doi{10.1080/00224065.1992.11979403}.
	
	\bibitem[Bissell(1990)]{bissell1990reliable}
	AF~Bissell.
	\newblock How reliable is your capability index?
	\newblock \emph{Journal of the Royal Statistical Society Series C: Applied
		Statistics}, 39\penalty0 (3):\penalty0 331--340, 1990.
	
	\bibitem[Mahmoud et~al.(2010)Mahmoud, Henderson, Epprecht, and
	Woodall]{mahmoud2010estimating}
	Mahmoud~A. Mahmoud, G.~Robin Henderson, Eugenio~K. Epprecht, and William~H.
	Woodall.
	\newblock Estimating the {{Standard Deviation}} in {{Quality-Control
			Applications}}.
	\newblock \emph{Journal of Quality Technology}, 42\penalty0 (4):\penalty0
	348--357, October 2010.
	\newblock ISSN 0022-4065, 2575-6230.
	\newblock \doi{10.1080/00224065.2010.11917832}.
	
	\bibitem[Chen and Pearn(1997)]{chen1997application}
	K.~S. Chen and W.~L. Pearn.
	\newblock An application of non-normal process capability indices.
	\newblock \emph{Quality and Reliability Engineering International}, 13\penalty0
	(6):\penalty0 355--360, 1997.
	\newblock ISSN 1099-1638.
	\newblock
	\doi{10.1002/(SICI)1099-1638(199711/12)13:6<355::AID-QRE125>3.0.CO;2-V}.
	
	\bibitem[Clements(1989)]{clements1989process}
	John~A Clements.
	\newblock Process capability calculations for non-normal distributions.
	\newblock \emph{Quality progress}, 22:\penalty0 95--100, 1989.
	
	\bibitem[Chen and Pearn(2001)]{chen2001capability}
	Kuen-Suan Chen and Wen-Lee Pearn.
	\newblock Capability indices for processes with asymmetric tolerances.
	\newblock \emph{Journal of the Chinese Institute of Engineers}, 24\penalty0
	(5):\penalty0 559--568, July 2001.
	\newblock ISSN 0253-3839, 2158-7299.
	\newblock \doi{10.1080/02533839.2001.9670652}.
	
	\bibitem[Abbasi~Ganji and Sadeghpour~Gildeh(2016)]{abbasi2016class}
	Z.~Abbasi~Ganji and B.~Sadeghpour~Gildeh.
	\newblock A class of process capability indices for asymmetric tolerances.
	\newblock \emph{Quality Engineering}, 28\penalty0 (4):\penalty0 441--454,
	October 2016.
	\newblock ISSN 0898-2112, 1532-4222.
	\newblock \doi{10.1080/08982112.2016.1168524}.
	
	\bibitem[Chan et~al.(1988)Chan, Cheng, and Spiring]{chan1988new}
	Lai~K. Chan, Smiley~W. Cheng, and Frederick~A. Spiring.
	\newblock A {{New Measure}} of {{Process Capability}}: {\emph{
	}}{{{\emph{C}}}}{\emph{{\textsubscript{pm}}}}{\emph{ }}.
	\newblock \emph{Journal of Quality Technology}, 20\penalty0 (3):\penalty0
	162--175, July 1988.
	\newblock ISSN 0022-4065, 2575-6230.
	\newblock \doi{10.1080/00224065.1988.11979102}.
	
	\bibitem[Boyles(1991)]{boyles1991taguchi}
	Russell~A. Boyles.
	\newblock The {{Taguchi Capability Index}}.
	\newblock \emph{Journal of Quality Technology}, 23\penalty0 (1):\penalty0
	17--26, January 1991.
	\newblock ISSN 0022-4065, 2575-6230.
	\newblock \doi{10.1080/00224065.1991.11979279}.
	
	\bibitem[V{\"a}nnman(1995)]{vannman1995unified}
	Kerstin V{\"a}nnman.
	\newblock A unified approach to capability indices.
	\newblock \emph{Statistica Sinica}, pages 805--820, 1995.
	
	\bibitem[Jiang and Yang(2026{\natexlab{b}})]{jiang2026finite}
	Fei Jiang and Lei Yang.
	\newblock Finite-sample decision instability in threshold-based process
	capability approval.
	\newblock \emph{arXiv:2603.11315}, 2026{\natexlab{b}}.
	
	\bibitem[Pendrill(2014)]{pendrill2014using}
	Leslie~R Pendrill.
	\newblock Using measurement uncertainty in decision-making and conformity
	assessment.
	\newblock \emph{Metrologia}, 51\penalty0 (4):\penalty0 S206--S218, 2014.
	
	\bibitem[ISO(ISO 14253-1:2013 (2013))]{ISO14253-1-2013}
	ISO.
	\newblock Geometrical product specifications (gps) -- inspection by measurement
	of workpieces and measuring equipment -- part 1: Decision rules for proving
	conformity or nonconformity with specifications.
	\newblock \emph{International Organization for Standardization}, ISO
	14253-1:2013 (2013).
	
	\bibitem[Desimoni and Brunetti(2011)]{desimoni2011uncertainty}
	Elio Desimoni and Barbara Brunetti.
	\newblock Uncertainty of measurement and conformity assessment: a review.
	\newblock \emph{Analytical and Bioanalytical Chemistry}, 400\penalty0
	(6):\penalty0 1729--1741, 2011.
	
	\bibitem[Wald(1950)]{wald1950statistical}
	Abraham Wald.
	\newblock Statistical decision functions.
	\newblock In \emph{Breakthroughs in Statistics: Foundations and Basic Theory},
	pages 342--357. Springer, 1950.
	
	\bibitem[DeGroot(2005)]{degroot2005optimal}
	Morris~H DeGroot.
	\newblock \emph{Optimal statistical decisions}.
	\newblock John Wiley \& Sons, 2005.
	
	\bibitem[Berger(2013)]{berger2013statistical}
	James~O Berger.
	\newblock \emph{Statistical decision theory and Bayesian analysis}.
	\newblock Springer Science \& Business Media, 2013.
	
	\bibitem[Hosmer~Jr et~al.(2013)Hosmer~Jr, Lemeshow, and
	Sturdivant]{hosmer2013applied}
	David~W Hosmer~Jr, Stanley Lemeshow, and Rodney~X Sturdivant.
	\newblock \emph{Applied logistic regression}.
	\newblock John Wiley \& Sons, 2013.
	
	\bibitem[Chen and Guestrin(2016)]{chen2016xgboost}
	Tianqi Chen and Carlos Guestrin.
	\newblock Xgboost: A scalable tree boosting system.
	\newblock In \emph{Proceedings of the 22nd acm sigkdd international conference
		on knowledge discovery and data mining}, pages 785--794, 2016.
	
	\bibitem[Gneiting and Raftery(2007)]{gneiting2007strictly}
	Tilmann Gneiting and Adrian~E Raftery.
	\newblock Strictly proper scoring rules, prediction, and estimation.
	\newblock \emph{Journal of the American statistical Association}, 102\penalty0
	(477):\penalty0 359--378, 2007.
	
	\bibitem[Guo et~al.(2017)Guo, Pleiss, Sun, and Weinberger]{guo2017calibration}
	Chuan Guo, Geoff Pleiss, Yu~Sun, and Kilian~Q Weinberger.
	\newblock On calibration of modern neural networks.
	\newblock In \emph{International conference on machine learning}, pages
	1321--1330. PMLR, 2017.
	
	\bibitem[Senge et~al.(2014)Senge, B{\"o}sner, Dembczy{\'n}ski, Haasenritter,
	Hirsch, Donner-Banzhoff, and H{\"u}llermeier]{senge2014reliable}
	Robin Senge, Stefan B{\"o}sner, Krzysztof Dembczy{\'n}ski, J{\"o}rg
	Haasenritter, Oliver Hirsch, Norbert Donner-Banzhoff, and Eyke
	H{\"u}llermeier.
	\newblock Reliable classification: Learning classifiers that distinguish
	aleatoric and epistemic uncertainty.
	\newblock \emph{Information Sciences}, 255:\penalty0 16--29, 2014.
	
	\bibitem[Jiang and Yang(2026{\natexlab{c}})]{jiang2026risk}
	Fei Jiang and Lei Yang.
	\newblock Risk-calibrated process capability approval with finite samples.
	\newblock \emph{arXiv preprint arXiv:2603.14479}, 2026{\natexlab{c}}.
	
	\bibitem[{Joint Committee for Guides in Metrology (JCGM)}(2012)]{jcgm106}
	{Joint Committee for Guides in Metrology (JCGM)}.
	\newblock Evaluation of measurement data --- the role of measurement
	uncertainty in conformity assessment, 2012.
	\newblock URL
	\url{https://www.bipm.org/documents/20126/2071204/JCGM_106_2012_E.pdf}.
	\newblock JCGM 106:2012.
	
	\bibitem[{International Organization for Standardization
		(ISO)}(2017)]{iso17025}
	{International Organization for Standardization (ISO)}.
	\newblock Iso/iec 17025:2017 --- general requirements for the competence of
	testing and calibration laboratories, 2017.
	\newblock URL \url{https://www.iso.org/standard/66912.html}.
	\newblock ISO/IEC 17025:2017.
	
	\bibitem[Van~der Vaart(2000)]{van2000asymptotic}
	Aad~W Van~der Vaart.
	\newblock \emph{Asymptotic statistics}, volume~3.
	\newblock Cambridge university press, 2000.
	
	\bibitem[Serfling(2009)]{serfling2009approximation}
	Robert~J Serfling.
	\newblock \emph{Approximation theorems of mathematical statistics}.
	\newblock John Wiley \& Sons, 2009.
	
	\bibitem[Deleryd(1998)]{deleryd1998gap}
	Mats Deleryd.
	\newblock On the gap between theory and practice of process capability studies.
	\newblock \emph{International Journal of Quality \& Reliability Management},
	15\penalty0 (2):\penalty0 178--191, 1998.
	
	\bibitem[Lehmann and Casella(1998)]{lehmann1998theory}
	Erich~Leo Lehmann and George Casella.
	\newblock \emph{Theory of point estimation}.
	\newblock Springer, 1998.
	
	\bibitem[Casella and Berger(2024)]{casella2024statistical}
	George Casella and Roger Berger.
	\newblock \emph{Statistical inference}.
	\newblock Chapman and Hall/CRC, 2024.
	
	\bibitem[Hand(2006)]{hand2006classifier}
	David~J Hand.
	\newblock Classifier technology and the illusion of progress.
	\newblock 2006.
	
	\bibitem[Breiman(1996)]{breiman1996stacked}
	Leo Breiman.
	\newblock Stacked regressions.
	\newblock \emph{Machine learning}, 24\penalty0 (1):\penalty0 49--64, 1996.
	
	\bibitem[Hastie(2017)]{hastie2017generalized}
	Trevor~J Hastie.
	\newblock Generalized additive models.
	\newblock \emph{Statistical models in S}, pages 249--307, 2017.
	
	\bibitem[{AIAG}(2010)]{aiag2010msa}
	{AIAG}.
	\newblock \emph{Measurement Systems Analysis (MSA) Reference Manual}.
	\newblock Automotive Industry Action Group, Southfield, MI, 4th edition, 2010.
	
	\bibitem[Shao and Tu(2012)]{shao2012jackknife}
	Jun Shao and Dongsheng Tu.
	\newblock \emph{The jackknife and bootstrap}.
	\newblock Springer Science \& Business Media, 2012.
	
\end{thebibliography}

\end{document}